\documentclass[a4paper,fleqn]{cas-sc}

\usepackage[numbers]{natbib}

\usepackage{amssymb}
\usepackage{amsmath,amsthm}
\usepackage[noend]{algpseudocode}
\usepackage{algorithm} 
\usepackage{array}
\usepackage[caption=false,font=footnotesize]{subfig}
\usepackage{stfloats}
\usepackage{url}
\usepackage{cases}
\usepackage{stfloats}
\usepackage{upgreek}
\usepackage{balance}
\usepackage{color}
\usepackage[dvipsnames]{xcolor}
\usepackage{verbatim}

\usepackage{graphicx}

\usepackage[T1]{fontenc} 

\usepackage{threeparttable}
\usepackage{multirow}

\usepackage{siunitx}
\usepackage{xcolor}
\usepackage[normalem]{ulem}
\usepackage{mleftright}
\usepackage[]{footmisc}
\mleftright

\DeclareMathOperator{\trace}{Tr}
\DeclareMathOperator{\diag}{\mathrm{diag}}
\newcommand{\paren}[1]{\left({#1}\right)}
\newcommand{\bracket}[1]{{\left [{#1}\right ]}}
\newcommand{\braces}[1]{{\left\{ {#1}\right\}}} 
\newcommand{\ith}[1]    {{#1}{\text{-th}}}
\newcommand{\rr}{_\mathrm{r}}
\newcommand{\cc}{_\mathrm{c}}

\newcommand{\B}{\textrm{B}}
\newcommand{\rnr}{_{\mathrm{r},n_\mathrm{r}}}
\newcommand{\target}{\mathrm{t}}
\newcommand{\MM}{\mathit{M}}

\newcommand{\dui}{\mathbf{d}_{\mathrm{u},i}\bracket{k,l}}

\newcommand{\PiB}{\mathbf{P}_{\textrm{u},i}\bracket{k}}
\newcommand{\PiBH}{\mathbf{P}^\dagger_{\textrm{u},i}\bracket{k}}

\newcommand{\PBj}{\mathbf{P}_{\textrm{d},j}\bracket{k}}

\newcommand{\PBjH}{\mathbf{P}^\dagger_{\textrm{d},j}\bracket{k}}
\newcommand{\PBg}{\mathbf{P}_{\textrm{d},g}\bracket{k}}

\newcommand{\UiB}{\mathbf{U}_{\textrm{u},i}\bracket{k}}

\newcommand{\UiBH}{\mathbf{U}^\dagger_{\textrm{u},i}\bracket{k}}

\newcommand{\WiB}{\mathbf{W}_{\textrm{u},i}\bracket{k}}

\newcommand{\UBj}{\mathbf{U}_{\textrm{d},j}\bracket{k}}

\newcommand{\UBjH}{\mathbf{U}^\dagger_{\textrm{d},j}\bracket{k}}

\newcommand{\WBj}{\mathbf{W}_{\mathrm{d},j}\bracket{k}}

\newcommand{\Wrnr}{\mathbf{W}_{\mathrm{r},n_\mathrm{r}}}

\newcommand{\HrB}{\mathbf{H}_{\textrm{rB}}}

\newcommand{\HBj}{\mathbf{H}_{\textrm{B},j}}

\newcommand{\HBB}{\mathbf{H}_{\mathrm{BB}}}

\newcommand{\HiB}{\mathbf{H}_{i,\textrm{B}}}
\newcommand{\HiBH}{\mathbf{H}^\dagger_{i,\textrm{B}}}

\newcommand{\Hij}{\mathbf{H}_{i,j}}

\newcommand{\sfrac}[2]{#1/#2}

\theoremstyle{definition}

\begin{document}

\shorttitle{Distributed MRMC - I}


\title [mode = title]{Co-Designing Statistical MIMO Radar and In-band Full-Duplex Multi-User MIMO Communications -- Part I: Signal Processing}    

\tnotetext[1]{K. V. M. acknowledges support from the National Academies of Sciences, Engineering, and Medicine via the Army Research Laboratory Harry Diamond Distinguished Fellowship. The following co-authors are a member of EURASIP: Kumar Vijay Mishra.}

\author[label1]{Jiawei~Liu}
\author[label2]{Kumar Vijay Mishra}[orcid=0000-0002-5386-609X]
\author[label1]{Mohammad~Saquib}

\affiliation[label1]{organization={The University of Texas at Dallas},
            city={Richardson},
            postcode={TX 75080}, 
            country={USA}}
\affiliation[label2]{organization={United States DEVCOM Army Research Laboratory},
            city={Adelphi},
            postcode={MD 20783}, 
            country={USA}}

	\maketitle
\begin{abstract}
We consider a spectral sharing problem in which a statistical (or widely distributed) multiple-input multiple-output (MIMO) radar and an in-band full-duplex (IBFD) multi-user MIMO (MU-MIMO) communications system concurrently operate within the same frequency band. Prior works on joint MIMO-radar-MIMO-communications (MRMC) systems largely focus on either colocated MIMO radars, half-duplex MIMO communications, single-user scenarios, omit practical constraints (clutter, uplink [UL]/downlink [DL] transmit powers, UL/DL quality-of-service, and peak-to-average-power ratio), or MRMC co-existence that employs separate transmit/receive units. The purpose of this and companion papers (Part II and III) is to co-design an MRMC framework that addresses all of these issues. In this paper, we propose signal processing for a distributed IBFD MRMC, where radar receiver is designed to additionally exploit the downlink communications signals reflected from a radar target. Extensive numerical experiments show that our methods improve radar target detection over conventional codes and yield a higher achievable data rate than standard precoders. The following companion paper (Part II) describes the theory and procedure of our algorithm to solve the non-convex design problem. The final companion paper (Part II) considers the case of multiple targets and examines the tracking performance of our MRMC system. 
\end{abstract}


\section{Introduction}
\color{black}Severe crowding of the electromagnetic spectrum in recent years has led to complex challenges in designing radar and communications systems that operate in the same bands \cite{mishra2019toward}.  
Both systems need wide bandwidth to provide a designated quality-of-service (QoS). Sometimes, their typical bandwidths may not have the same order of magnitude and they may operate with only a partial spectral overlap. Whereas a high-resolution detection of radar targets requires significant transmit signal bandwidths\cite{skolnik2008radar}, 
the wireless cellular networks need access to a broad spectrum to support high data rates \cite{Multiuser,cover2006elements}.
With the rapid surge in mobile data traffic, network operators worldwide have turned to a higher frequency spectrum to accommodate the rise in data usage \cite{mishra2019toward}. In addition, the continuous scaling up of the carrier frequencies and deployment of wireless communications have propelled the spectrum regulators to grant civilian communications systems access to the spectrum traditionally reserved for radar/sensing applications. This policy shift has sparked the trend of coexisting and even converging the radar and communications functions\cite{mishra2019toward}. 

\color{black} 
Broadly, the two most common approaches are sharing with spectral overlap (or \textit{co-existence}) and functional spectrum-sharing (or \textit{co-design}) \cite{mishra2019toward}. \textcolor{black}{In the former, the transmitter (Tx) and receiver (Rx) of radar and communications are separate units that operate within the same spectrum using different waveforms \cite{interferencealignment,ayyar2019robust}.} 
The latter combines the two systems at either Tx, Rx, or both in a single hardware platform and employs a common waveform
\cite{dokhanchi2019mmwave,duggal2020doppler}. These spectrum-sharing solutions also depend on the level of cooperation between radar and communications. In a \textit{selfish} paradigm, the overall architecture usually promotes the performance of only one system leading to radar-centric \cite{alaee2019discrete,bao2019precoding,slavik2019cognitive} and communications-centric \cite{ayyar2019robust} co-existence solutions. On the other hand, the \textit{holistic} solution relies on extensive cooperation between the two systems in transmitting strategies and receiving processing \cite{mahal2017spectral,MCMIMO_RadComm,qian2018joint,rihan2018optimum,Lops2019serveillance,biswas2018fdqos,he2019performance}. The exchange of information, such as the channel state information (CSI), may also be facilitated via a fusion center \cite{MCMIMO_RadComm,he2019performance}.
The spectral cooperation enables both systems to benefit from increased degrees of freedom (DoFs) and allows joint optimization of system parameters through one \cite{MCMIMO_RadComm,qian2018joint} or more \cite{biswas2018fdqos,dokhanchi2020multi} objective functions. For example, the communications signals decoded at the radar Rx may be used to enhance target detection/localization \cite{biswas2018fdqos,he2019performance}. Similarly, communications Rxs may improve error rates by extracting symbols embedded in the echoes reflected off the radar targets \cite{mishra2019toward}. In this paper, we focus on the holistic spectral co-design.

The approaches above do not readily extend to multiple-input multiple-output (MIMO) configuration, which employs several Tx and Rx antennas to achieve high spectrum efficiency \cite{Multiuser,haimovich2008mimo}. MIMO configuration enhances communications capacity, provides spatial diversity, and exploits multi-path propagation \cite{Multiuser}. \textcolor{black}{Further, recent developments in massive MIMO \cite{Lops2020uplink,mishra2019toward} utilize uplink/downlink (UL/DL) channel reciprocity via a vast number of service antennas to serve a lower number of mobile users with time-division-duplexing.}

\color{red}
\begin{table}
\centering
\caption{Comparison with the state-of-the-art}
\label{tbl:priorcomp}
\resizebox{\textwidth}{!}{%
\begin{threeparttable}
\begin{tabular}{l||c|c|c||c|c|c||c|l}
\hline \multirow{2}{*}{cf.} & \multicolumn{3}{c||}{Radar}   & \multicolumn{3}{c||}{Communications} & \multirow{2}{*}{Design metric} & \multirow{2}{*}{Design objective}  \\\cline{2-7}
                         & Model & Targets & Clutter & Model & Duplexing & Users & &   \\
\hline \hline \cite{MCMIMO_RadComm}  & C-MIMO\tnote{a} & Static & Yes & P2P MIMO\tnote{b} & HD & SU & Radar SINR & Joint waveforms\\
\hline \cite{he2019performance} & D-MIMO\tnote{c} & Static & Yes & D-MIMO & HD & SU  & n/a\tnote{d} & Target localization 
\\
\hline \cite{Lops2020uplink}  & Monostatic & Static & Yes & M-MIMO\tnote{e} & HD (UL) & SU  & MI & Joint Rx filters, BFs\\
\hline \cite{liu2018mimo}  & C-MIMO & Static & No & MIMO & HD (DL)  & MU  & Transmit power & Joint transmit BFs\\
\hline \cite{biswas2018fdqos}  & C-MIMO & Moving & No & MIMO & FD & MU & Radar $P_d$ & BFs, radar waveform \\
\hline \cite{Valkama2021beamformer}  & C-MIMO & Moving & No & MIMO & FD & MU & Beamformer power & Joint BF \\
\hline \cite{Smida2022ISAC} & C-MIMO & Moving & No & MIMO & IBFD & MU & Total SNR & Joint BF-SIC design \\
\hline \cite{Zeng2022WaveFDISAC} & Monostatic & Moving & No & P2P SISO & IBFD & SU & Mutual interference &  Waveform \\
\hline This paper & D-MIMO & Moving & Yes & MIMO & IBFD & MU & CWSM & Joint waveform-precoders-filter \\
\hline
\end{tabular}
\begin{tablenotes}[para]
\item[a] C-MIMO: Colocated MIMO \item[b] P2P: Point-to-point \item[c] D-MIMO: Distributed MIMO \item[d] n/a: Not applicable \item[e] M-MIMO: Massive MIMO
\end{tablenotes}
\end{threeparttable}
}
\end{table}
\normalsize
\color{black}

Similarly, MIMO radars outweigh an equivalent phased array radar by offering higher angular resolution with fewer antennas, spatial diversity, and improved parameter identifiability by exploiting waveform diversity \cite{haimovich2008mimo}. 
In a \textit{colocated} MIMO radar \cite{mishra2019cognitive}, the radar cross-section is identical to closely-spaced antennas. On the contrary, the antennas in a \textit{widely distributed} MIMO radar are sufficiently separated from each other such that the same target projects a different radar cross-section to each Tx-Rx pair; this spatial diversity is advantageous in detecting targets with small backscatters and low speed \cite{hongbin_movingtarget,sun2019target,sun2024widely}. The distributed system is also termed a \textit{statistical} MIMO radar because the path gain vectors in a distributed array are modeled as independent statistical variables \cite{hongbin_movingtarget,Jammer_game,NaghshTSP2017,sun2019target}. 

The increased DoFs, sharing of antennas, and higher dimensional optimization exacerbate spectrum sharing in a joint MIMO-radar-MIMO-communications (MRMC) architecture \cite{alaee2020information,dokhanchi2020multi}. 
Early works on MRMC proposed null space projection beamforming, which projects the colocated MIMO radar signals onto the null space of the interference channel matrix from radar Tx to MIMO communications Rx \cite{khawar2015target}. The MRMC processing techniques include matrix completion 
\cite{MCMIMO_RadComm}, single base station (BS) interference mitigation \cite{khawar2015target}, and switched small singular value space projection 
\cite{mahal2017spectral}.
Among waveforms, \cite{bao2019precoding} analyzes orthogonal frequency-division multiplexing for a MIMO radar to coexist with a communications system. On the other hand, \cite{qian2018joint,rihan2018optimum} suggest optimal space-time transmit waveforms for a colocated MIMO radar co-designed with a point-to-point MIMO communications codebook.
	
Nearly all of these works focus on single-user  MIMO communications and colocated MIMO radars. To generalize MRMC, \cite{cheng2019miso} investigated a novel constructive-interference-based precoding optimization for colocated MIMO radar and DL MU multiple-input single-output communications. This was later extended to the co-existence of MIMO radar with MU-MIMO communications \cite{biswas2018fdqos} through multiple radar transmit beamforming approaches that keep the original modulation and communications data rate unaffected. In quite a few recent studies \cite{Lops2020uplink,Lops2019serveillance}, either radar or communications is in MIMO configuration.

Co-design with statistical MIMO radar remains relatively unexamined in these prior works. The model in \cite{he2019performance} comprises widely distributed MIMO radar but studies co-existence with simplistic point-2-point MIMO communications. The distributed radar proposed in \cite{sedighi2021localization} exploits a communications waveform but operates in passive (receive-only) mode. The MIMO communications model in the studies above is limited to half-duplex (HD) rather than full-duplex (FD), which allows concurrent transmission and reception, usually in non-overlapping frequencies. The throughput is further doubled through in-band FD (IBFD) communications that enable UL and DL to function in a single time/frequency channel through advanced self-interference (SI) cancellation techniques \cite{Barneto2021FDcommsensing,Hassani2020IBFD,FD_WMMSE}. 
The IBFD technology has been recently explored for joint radar-communications systems to facilitate communications transmission while also receiving the target echoes \cite{Barneto2021FDcommsensing}. In \cite{biswas2018fdqos}, the BS operates in MU IBFD mode and simultaneously serves multiple DL and UL user equipment (UEs) while coexisting with a colocated MIMO radar. For a co-existence scenario with a colocated MIMO radar, another study in \cite{Singh2020FDRadar} considers the precoder design of an FD cellular system given imperfect channel state information and hardware impairments. Furthermore, \cite{Hassani2020IBFD} prototyped an IBFD communications system integrated with a monostatic Doppler radar. However, these IBFD MRMC studies do not consider statistical MIMO radar.

The performance metrics to design radar and communications systems are not identical because of different system goals. In general, a communications system strives to achieve high data rates while a radar performs detection, estimation, and tracking.  The MI is a well-studied metric in MU-MIMO communications for transmit precoder design \cite{Luo2011IterativeWMMSE,MSE_FD}. Several recent works on radar code design adopt the MI as a design criterion \cite{Colornoise_waveform,Jammer_game,NaghshTSP2017,20MISTAP}. In particular, \cite{Colornoise_waveform,NaghshTSP2017} show that maximizing the MI between the radar receive signal and the target response leads to a better detection performance in the presence of the Gaussian noise. For the radar-communications co-design, \cite{Xiaodong_Overlaid} proposed an overall metric for the joint radar-communications system dubbed the compound rate, a combination of a radar signal-to-interference-plus-noise ratio and a point-to-point communications achievable rate. Some recent works \cite{alaee2020information,dokhanchi2020multi} suggest mutual information (MI) as a common performance metric for the joint radar and communications system. However, there is still a lack of effort in applying this metric to the MIMO radar and MU-MIMO communications co-design problems.  

This work addresses these gaps by proposing a co-design scheme for a statistical MIMO radar and an IBFD MU-MIMO communications system. As in most IBFD MU-MIMO systems \cite{biswas2018fdqos,Singh2020FDRadar,FD_WMMSE}, the BS operates in FD while all the DL or UL UEs is in HD mode. Further, unlike many prior works that focus solely on one specific system goal and often in isolation with other processing modules, we jointly design the UL/DL precoders, MIMO radar waveform matrix, and linear receive filters (LRFs) for both systems. Our novel approach exploits mutual information (MI) by proposing a novel \textit{compounded-and-weighted sum MI} (CWSM) to measure the combined performance of the radar and communications systems. In contrast to prior works in \cite{biswas2018fdqos,he2019performance} that used information-theoretic metrics for only communications, we introduce a common MI-based criterion to enable a joint MRMC design. Table~\ref{tbl:priorcomp} summarizes the differences in the model and methods of prior art and this work.

Our co-design also accounts for several practical constraints, including the maximum UL/DL transmit powers, the QoS of the UL/DL quantified by their respective minimum achievable rates, and the peak-to-average-power-ratio (PAR) of the MIMO radar waveform. It is common among communications literature to identify a UL/DL UE's QoS with its minimum achievable rate \cite{MIMOCOMSecrecy,biswas2018fdqos}. Adopting low PAR waveforms is crucial for achieving energy- and cost-efficient RF front-ends \cite{NaghshTSP2017}. However, prior works have not considered the QoS and the PAR constraints concurrently for a joint statistical MIMO radar and MU-MIMO FD communications system design. We herein address the non-convex CWSM maximization problem's challenges subject to non-convex constraints, namely the QoS and the PAR constraints. To this end, in the following companion paper (Part II) \cite{liu2024codesigningpart2}, we develop an alternating algorithm that incorporates both the block coordinate descent (BCD) and the alternating projection (AP) methods. With our optimized radar codes, the probability of detection is enhanced up to $13$\% over conventional radar waveforms at a given false alarm rate. Furthermore, our optimized precoders yield up to a $30$\% higher rate than standard precoders. Preliminary results of this work appeared in our conference publication \cite{Liu2022precoder}, where only precoder design was considered. In the final companion paper (Part III) \cite{liu2024codesigningpart3}, we develop tracking algorithm for the case of multiple targets in a joint statistical MIMO radar and IBFD MU-MIMO communications system.

The rest of the paper is organized as follows. The next section describes the system models of the statistical MIMO radar and the IBFD-MIMO communications system, respectively. We present the co-design-based receiver signal processing for the MIMO radar and IBFD MU-MIMO communications in Sections~\ref{sec:radar_total_receive} and \ref{sec:comm_rx}, respectively. We formulate the CWSM maximization problem in Section \ref{sec: formulation}. We then employ the BCD-AP MRMC procedure from the following companion paper (Part II) \cite{liu2024codesigningpart2} to solve the optimization problem iteratively and validate the proposed technique through numerical experiments in Section \ref{sec:numerical} before concluding in Section \ref{sec:conclusion}.
	
Throughout this paper, lowercase regular, lowercase boldface and uppercase boldface letters denote scalars, vectors and matrices, respectively. We use $I(\mathbf{X};\mathbf{Y})$ and $H\paren{\mathbf{X}|\mathbf{Y}}$ to denote, MI and conditional entropy between two random variables $\mathbf{X}$ and $\mathbf{Y}$, respectively. The notations $\mathbf{Y}\bracket{k}$, $\mathbf{y}\bracket{k}$, and $y\bracket{k}$ denote the value of time-variant matrix $\mathbf{Y}$, vector $\mathbf{y}$ and scalar $y$ at discrete-time index $k$, respectively; $\mathbf{1}_{N}$ is a vector of size $N$ with all ones; $\mathbb{C}$ and $\mathbb{R}$ represent sets of complex and real numbers, respectively; a circularly symmetric complex Gaussian (CSCG) vector $\mathbf{q}$ with $N$ elements and power spectral density $\mathcal{N}_0$ is $\mathbf{q}\sim\mathcal{CN}(0,\mathcal{N}_0\mathbf{I}_{N})$; $(\cdot)^{\star}$ is the solution of the optimization problem; $\mathbb{E}\bracket{\cdot}$ is the statistical expectation; $\trace\{\mathbf{R}\}$, $\mathbf{R}^\top$, $\mathbf{R}^\dagger$, $\mathbf{R}^\ast$, $\left| \mathbf{R}\right|$, $\mathbf{R}\succeq\mathbf{0}$, and $\mathbf{R}\paren{m,n}$ are the trace, transpose, Hermitian transpose, element-wise complex conjugate, determinant, positive semi-definiteness and $\ith{\paren{m,n}}$ entry of matrix $\mathbf{R}$, respectively; set $\mathbb{Z}_{+}(L)$ denotes $\left\lbrace1,\dots,L\right\rbrace$;  $\mathbf{x}\succeq\mathbf{y}$ denotes component-wise inequality between vectors $\mathbf{x}$ and $\mathbf{y}$; $x^+$ represents $\max(x,0)$; $x^{\paren{t}}\paren{\cdot}$ is the $\ith{t}$ iterate of an iterative function $x\paren{\cdot}$; $\inf(\cdot)$ is the infimum of its argument; $\odot$ denotes the Hadamard product; and $\oplus$ is the direct sum. 
\color{black} Table~\ref{table_parameter} summarizes the symbols used in this paper.

	\begin{table}[t] 
	\color{black}
		\renewcommand{\arraystretch}{1.3} 
		\caption{Glossary of Notations}
		\label{table_parameter}
		\centering
		\begin{tabular}{l||p{59mm}}
			\hline
			\bfseries Symbol & \bfseries Description\\
			\hline
			$\mathit{M}\rr$ ($N\rr$)& number of the MIMO radar Tx (Rx) antennas\\
			\hline
			$\mathit{M}\cc$ ($N\cc$)&number of the BS Tx (Rx) antennas\\
			\hline
			$\mathit{I}$ ($\mathit{J}$)& number of the UL (DL) UEs\\
			\hline
			$K$&number of radar pulses or communications frames 
			\\
			\hline
			$N^{\textrm{u}}_i$ ($N^{\textrm{d}}_{j}$)&number of antennas in the $\ith{i}$ UL ($\ith{j}$ DL) UE \\
			\hline
			$\mathrm{D}^\textrm{u}_{i}$ ($\mathrm{D}^\textrm{d}_{j}$)&number of the $\ith{i}$ UL ($\ith{j}$ DL) data streams \\
			\hline
			$\mathbf{d}_{\mathrm{u},i}$ ($\mathbf{d}_{\mathrm{d},j}$)& symbol vector of the $\ith{i}$ UL ($\ith{j}$ DL) UE  \\
			\hline
			$\mathbf{a}_{m\rr}$& radar codeword for the $\ith{m\rr}$ radar Tx\\
			\hline
			$\mathbf{A}$& total radar transmit waveform matrix\\
			\hline
			$\mathbf{h}_{\text{rt,}n\rr}$& target response vector to $\ith{n\rr}$ radar Rx\\
			\hline
			$\mathbf{h}_{\text{Bt,}n\rr}$&  channel for BS-to-target-to-$\ith{n\rr}$ radar Rx\\
			\hline
			$\HiB$ ($\HBj$)& channel matrix for $\ith{i}$ UL UE ($\ith{j}$ DL UE)\\
			\hline
			$\HBB$& FD self-interference matrix\\
			\hline
			$\Hij$& channel matrix from $\ith{i}$ UL UE to $\ith{j}$ DL UE\\
			\hline
			$\HrB$& channel matrix from the MIMO radar to the BS\\
			\hline
			$\PiB$ ($\PBj$)& precoder of $\ith{i}$ UL ($\ith{j}$ DL) UE for $\ith{k}$ frame\\
			\hline
			$\{\mathbf{P}\}$& $\braces{\PiB,\PBj,\forall i,j,k}$\\
			\hline
			$\mathbf{U}_{\mathrm{r},n\rr}$& the LRF filter at $\ith{n\rr}$ MIMO radar Rx\\
			\hline
			$\UiB$ ($\UBj$)& Rx of $\ith{i}$ UL ($\ith{j}$ DL) UE for $\ith{k}$ frame\\
			\hline
			$\{\mathbf{U}\}$& $\braces{\UiB, \UBj, \mathbf{U}\rnr, \forall i,j,k,n\rr}$\\
			\hline
			$\Wrnr$& weight matrix for $\ith{n\rr}$ MIMO radar Rx\\
			\hline
			$\WiB$ ($\WBj$)& weight matrix for $\ith{i}$ UL ($\ith{j}$ DL) UE\\
			\hline
		\end{tabular}
	\end{table}
\color{black}
\section{Spectral Co-Design System Model}
\label{sec:system}
 \begin{figure}[t]
	\centering
	\includegraphics[width=1.0\columnwidth]{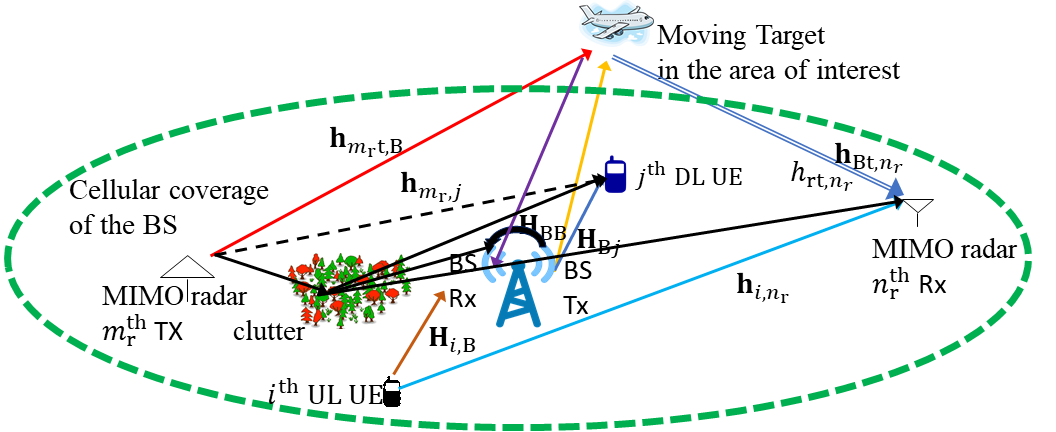}
	\caption{Co-design system model comprising a statistical (widely distributed) MIMO radar and IBFD MU-MIMO communications.}
	\label{fig:setup}
\end{figure}
\textcolor{black}{Consider a two-dimensional (2-D) $\left(x\textrm{-}y \right)$ Cartesian plane on which the $M_\mathrm{r}$ Txs and $N_\mathrm{r}$ Rxs of a statistical MIMO radar, the BS, $I$ UL UEs, and $J$ DL UEs of the IBFD MU-MIMO communications system are located at the coordinates $\left(x_{m_\mathrm{r}},y_{m_\mathrm{r}}\right)$, $\left(x_{n_\mathrm{r}},y_{n_\mathrm{r}} \right)$,  $\paren{x_{\mathrm{B}},y_{\mathrm{B}}}$, $\paren{x_{\textrm{UL},i},y_{\textrm{UL},i}}$, and $\paren{x_{\mathrm{DL,}j},y_{\textrm{DL},j}}$, respectively, for all $m_\mathrm{r}\in{Z}_{+}(M_\mathrm{r})$, $n_\mathrm{r}\in\mathbb{Z}_{+}\paren{N_\mathrm{r}}$, $i\in\mathbb{Z}_{+}\paren{I}$, and $j\in\mathbb{Z}_{+}\paren{J}$ (\figurename{~\ref{fig:setup}}).} The statistical MIMO radar operates within the same transmit spectrum as an IBFD MU-MIMO communications system. Here, the radar aims to detect a target moving within the cellular coverage of the BS. The communications system serves the UL/DL UEs with desired achievable rates in the presence of the radar echoes. 
\subsection{Transmit Signal}
Each radar Tx emits a train of $\mathit{K}$ pulses at a uniform pulse repetition interval (PRI) $T_{\mathrm{r}}$ or \textit{fast-time}; the total duration $KT_{\mathrm{r}}$ is the \textit{coherent processing interval} (CPI) or \textit{slow-time} (\figurename{\;\ref{fig:transmissionmodel}}) and $\mathit{K}$ is chosen to avoid range migration during the CPI \cite{skolnik2008radar}. At the same time,  the BS and each UL UE continuously transmit DL and UL symbols, respectively. The radar pulse width is $T_\mathrm{p}= T_\mathrm{r}/N$, where $N$ is the number of sampled range bins or cells in a PRI. The UL/DL frame duration $T_\mathrm{f}$ and the UL/DL symbol duration $T_{\mathrm{s}}$ equal radar PRI and radar pulse width $T_{\mathrm{p}}$, respectively; i.e., $T_{\mathrm{f}}=T_{\mathrm{r}}$ and $T_{\mathrm{s}}=T_{\mathrm{p}}$. This implies that the number of UL/DL frames transmitted in the scheduling window is also $\mathit{K}$ and the number of UL/DL symbols per frame is $\sfrac{T_{\mathrm{f}}}{T_{\mathrm{s}}}=N$.

\subsection{Synchronization}
\begin{figure}[!t]
		\centering
		\includegraphics[width=1.0\columnwidth]{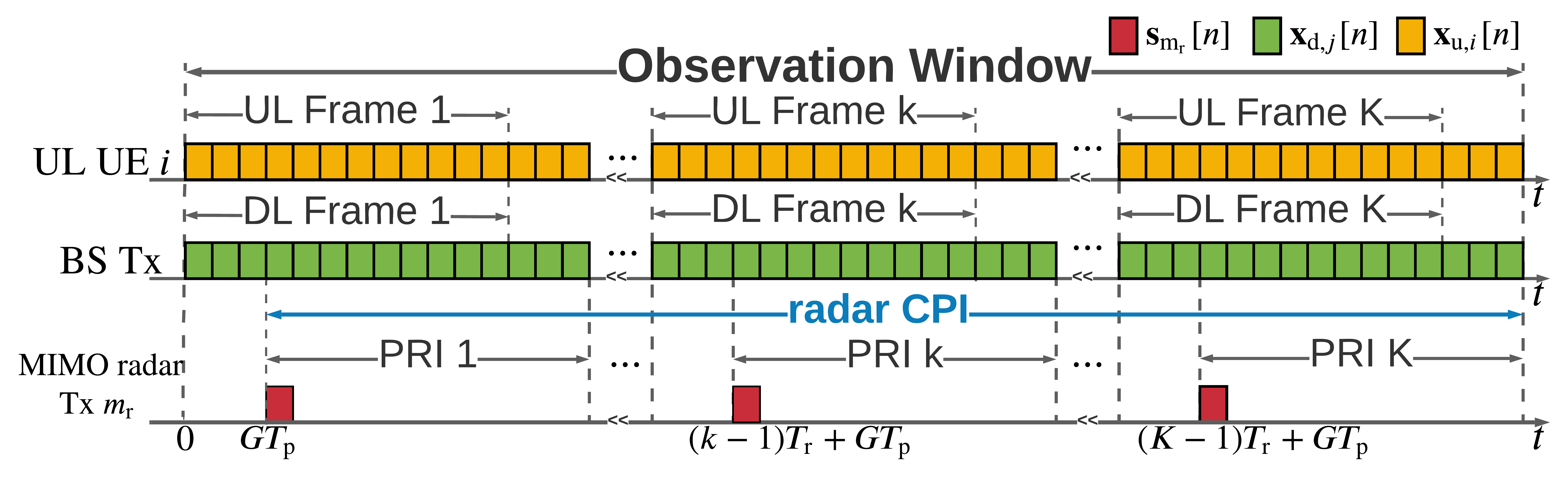}
		\caption{\textcolor{black}{Transmission sequence during the observation interval $t\in\bracket{0,KT_{\mathrm{r}}+GT_\mathrm{p}}$. Each bin represents a discrete-time radar/communications transmit signal of duration $T_{\mathrm{p}}$. The communications data transmissions occur continuously while pulsed radar Txs emit probing signals at the rate $1/T\rr$.}
		} 
		\label{fig:transmissionmodel}
	\end{figure}

The FD MU-MIMO communications system maintains the carrier and symbol synchronizations by periodically estimating the carrier frequency and phase \cite{Multiuser}. The Rxs of radar and communications employ the same sampling rates and, therefore, communications symbols and radar range cells are aligned in time \cite{rihan2018optimum,MCMIMO_RadComm}. The clocks at the BS and the MIMO radar are synchronized offline and periodically updated such that the clock offsets between the BS and MIMO radar Rxs are negligible \cite{interferencealignment}. Using the feedback of the BS via pilot symbols, radar Rxs can obtain the clock information of UL UEs. Note that this setup exploits the established clock synchronization standards that have been widely adopted in wireless communications and distributed sensing systems, e.g., the IEEE 1588 precision time protocol \cite{wang2020displaced, sedighi2021localization}. Note that our model is realistic and does not assume a fully synchronous transmission between the radar and communications, similar to the one in \cite{Lops2019serveillance}. As shown in \figurename{~\ref{fig:transmissionmodel}}, the $\ith{k}$ communications frame is transmitted at a duration of $GT_{\mathrm{p}}$, $G\in\mathbb{Z}\paren{N-1}$, before the $\ith{k}$ radar PRI. 
\color{black}
\subsubsection{Statistical MIMO Radar}
\label{sec: Radar_Tx_Signals}
Denote the \textit{narrowband} transmit pulse of the $m\rr$-th radar Tx by $\phi_{m_\mathrm{r}}\paren{t}$.  
The waveforms from all Txs form the waveform vector
\begin{align}
\boldsymbol{\phi}(t)=\left[ \phi_1(t),\dots,\phi_{M_\mathrm{r}}(t)\right]^\top\in\mathbb{C}^{M\rr},
\end{align}
and satisfy the orthonormality $\int_{T_\mathrm{p}}^{}\boldsymbol{\phi}(t)\boldsymbol{\phi}^\dagger(t)dt=\mathbf{I}_{M_\mathrm{r}}$. The radar code to modulate the pulse emitted by the $m\rr$ Tx in the $\ith{k}$ PRI is $a_{m\rr,k}$. During the observation window $t\in\bracket{0,KT\rr+GT_{\mathrm{p}}}$, the $\ith{m_\mathrm{r}}$ Tx emits the pulse train
\begin{align}
s_{m_\mathrm{r}}\paren{t}=\sum_{k=0}^{K-1}a_{m_\mathrm{r},k}\phi_{m_\mathrm{r}}\paren{t-kT\rr-GT_{\mathrm{p}}},\label{eq: radar_pulse_train}
\end{align}
where the support of $\phi_{m\rr}\paren{t}$ is $\left[0,T\rr\right)$ and, without loss of generality, $\phi_{m\rr}\paren{t}=\sqrt{\sfrac{1}{T_{\textrm{p}}}}e^{j2\pi\frac{m\rr}{T_{\textrm{p}}}t}$ for $m\rr\in\mathbb{Z}_+\paren{M\rr}$ for $t\in\left[0,T\rr\right)$. Define the radar code vector transmitted during the $\ith{k}$ PRI as $\mathbf{a}\bracket{k}=\bracket{a_{1,k},\cdots,a_{\mathit{M}\rr,k}}^\top\in\mathbb{C}^{M\rr}$ so that the MIMO radar code matrix is
\begin{align}
\mathbf{A}=\bracket{\mathbf{a}^\top\bracket{1};\cdots; \mathbf{a}^\top\bracket{\mathrm{\mathit{K}}}}=\bracket{\mathbf{a}_1,\cdots.\mathbf{a}_{M\rr}}\in\mathbb{C}^{\mathit{K}\times \mathit{M}\rr}. 
\end{align}
where $\mathbf{a}_{m\rr}\in\mathbb{C}^{K}$ is the code of the $\ith{m\rr}$ TX over all PRIs. The combined transmit signal vector is
\begin{align}
\mathbf{s}(t)=\bracket{s_1(t),\cdots,s_{M_\mathrm{r}}(t)}^\top\in\mathbb{C}^{M\rr}. 
\end{align}

\subsubsection{IBFD MU-MIMO Communications}
The BS and UEs operate in the FD and HD modes, respectively. During the observation window, the BS receives data frames from the $I$ UL UEs; concurrently, the $J$ DL UEs operating in the same band download data frames from the BS. The BS is equipped with $\mathit{M}_\mathrm{c}$ transmit and $N_{\mathrm{c}}$ receive antennas. The $i$-th UL UE and $j$-th DL UE  employ $N^{\textrm{u}}_{i}$ and $N^{\textrm{d}}_{j}$ transceive antennas, respectively. To achieve the maximum capacities of the UL and DL channels, number of BS Tx and Rx antennas are $\mathit{M}\cc\geq\sum_{j=1}^{\mathit{J}}N^{\textrm{d}}_{j}$ and $N\cc\geq\sum_{i=1}^{\mathit{I}}N^{\textrm{u}}_{i}$, respectively \cite{Multiuser}. 
A total of $\mathit{D}^{\textrm{u}}_{i}\leq N^{\textrm{u}}_{i}$ and $\mathit{D}^{\textrm{d}}_{j}\leq N^{\textrm{d}}_{j}$ unit-energy data streams are used by $i$-th UL UE and $j$-th DL UE, respectively. The symbol vectors sent by the $i$-th UL UE toward the BS and by the BS toward the $j$-th DL UE in the $\ith{l}$ symbol period of the $\ith{k}$ frame are $\mathbf{d}_{\mathrm{u},i}\bracket{k,l}\in \mathbb{C}^{D^\textrm{u}_{i}}$ and $\mathbf{d}_{\mathrm{d},j}\bracket{k,l}\in \mathbb{C}^{\mathit{D}^{\textrm{d}}_{j}}$, respectively; these are independent and identically distributed (i.i.d.) with $\mathbb{E}\bracket{\mathbf{d}_{\mathrm{d},j}\mathbf{d}^\dagger_{\mathrm{d},j}\bracket{k,l}}=\mathbb{E}\bracket{\mathbf{d}_{\mathrm{u},i}\mathbf{d}^\dagger_{\mathrm{u},i}\bracket{k,l}}=\mathbf{I}$ for $i\in\mathbb{Z}_+\braces{\mathit{I}}$, $k\in\mathbb{Z}_+\braces{\mathit{K}}$, and $l\in\mathbb{Z}_+\braces{N}$.

Denote the precoders for the $i$-th UL UE and the $j$-th DL UE at the $\ith{k}$ frame as $\PiB\in\mathbb{C}^{N^{\textrm{u}}_{i}\times \mathit{D}^{\textrm{u}}_{i}}$ and $\PBj\in\mathbb{C}^{\mathit{M}\cc\times \mathit{D}^{\textrm{d}}_{j}}$, respectively. The precoded transmit signal vectors for the $i$-th UL UE and $j$-th DL UE 
become 
\begin{align}
\mathbf{s}_{\textrm{u},i}\bracket{k,l}=\PiB\mathbf{d}_{\mathrm{u},i}\bracket{k,l}, 
\end{align}
and 
\begin{align}
\mathbf{s}_{\textrm{d},j}\bracket{k,l}=\PBj\mathbf{d}_{\mathrm{d},j}\bracket{k,l},
\end{align}
respectively. The total DL symbol vector broadcast by the BS in the same symbol period is 
\begin{align}
\mathbf{s}_{\mathrm{B}}\bracket{k,l}=\sum_{j=1}^{J}\mathbf{s}_{\textrm{d},j}\bracket{k,l}.
\end{align}
The transmit pulse shaping function used by the IBFD communications is $p_{\mathrm{T}}\paren{t}$. The transmit signals of $i$-th UL UE and BS are
\begin{flalign}
\mathbf{x}_{\mathrm{u},i}\paren{t}&=\sum_{k=0}^{K-1}\sum_{l=0}^{N-1}\mathbf{s}_{\mathrm{u},i}\bracket{k,l}p_{\mathrm{T}}\paren{t-(kN+l)T_{\mathrm{p}}}, \end{flalign}
and
\begin{flalign}
\mathbf{x}_{\mathrm{B}}\paren{t}&=\sum_{k=0}^{K-1}\sum_{l=0}^{N-1}\mathbf{s}_{\mathrm{B}}\bracket{k,l}p_{\mathrm{T}}\paren{t-\paren{k N+l}T_{\mathrm{p}}}.
\end{flalign}

\subsection{Channel}
For narrowband signaling, assume a block fading model for communications and radar channel gains, both of which remain constant during the observation window \cite{interferencealignment,mishra2019toward}. 
\subsubsection{Statistical MIMO Radar}\label{subsubsec:radar_channel}
The complex reflectivity of the target associated with the $\ith{m\rr}$ Tx - $\ith{n\rr}$ Rx path is modeled as a CSCG random variable $\alpha_{m\rr \target n\rr }\sim\mathcal{CN}(0,\eta^2_{m\rr\target n\rr})$, where $\eta^2_{m\rr\textrm{t}n\rr}$ denotes the average reflection power of the point target proportional to the target RCS\cite{NaghshTSP2017}; it remains constant over the CPI as per the Swerling I (block fading) target model \cite{haimovich2008mimo}.

Denote the velocity vector of the target by $\boldsymbol{\nu}_{t}\triangleq\left(\nu_\mathrm{\mathrm{x},t},\nu_\mathrm{\mathrm{y},t} \right)$, where $\nu_\mathrm{\mathrm{x},t}$ and $\nu_\mathrm{\mathrm{y},t}$ are deterministic, unknown $x$- and $y$-direction velocity components in the Cartesian plane.  
The Doppler frequency w.r.t. the $\ith{m\rr}$ Tx - $\ith{n\rr}$ Rx pair is \cite{hongbin_movingtarget} 
\begin{align}
f_{m_\mathrm{r}\target n_\mathrm{r}} = \frac{\nu_\mathrm{\mathrm{x},t}}{\lambda}(\cos\theta_{m_\mathrm{r}\target}+\cos\phi_{n_\mathrm{r}\target})+\frac{\nu_\mathrm{\mathrm{y},\target}}{\lambda}(\sin\theta_{m_\mathrm{r}\target}+\sin\phi_{n_\mathrm{r}\target}),
\end{align}\normalsize
where $\lambda$ is the carrier wavelength, $\theta_{m_\mathrm{r}}$ ($\phi_{n_\mathrm{r}}$) is the angle-of-departure (angle-of-arrival) or AoD (AoA) at the $\ith{m_\mathrm{r}}$ Tx ($\ith{n_\mathrm{r}}$ Rx). 
The narrowband assumption of $s_{m\rr}(t)$ allows approximation of the propagation delay arising from the reflection off an arbitrary scatterer in the $\ith{n_\mathrm{r}}$ target to that from its center of gravity, 
for all $m_\mathrm{r}\in\mathbb{Z}_{+}(M_\mathrm{r})$ and $n_\mathrm{r}\in\mathbb{Z}_{+}(N_\mathrm{r})$ \cite{haimovich2008mimo}. If center of gravity of the target is located at $(x_{\mathrm{t}},y_{\mathrm{t}})$, then the propagation delay w.r.t. $\ith{m\rr}$ Tx - $\ith{n\rr}$ Rx is 
\begin{align}
\zeta_{m\rr \target n\rr}
=\frac{\sqrt{\paren{x_{m\rr}-x_{t}}^2+\paren{y_{m\rr}-y_{t}}^2}}{c}
		+\frac{\sqrt{\paren{x_{n\rr}-x_{t}}^2+\paren{y_{n\rr}-y_{t}}^2}}{c},
\end{align}	\normalsize
where $c$ is the speed of the light. The slow-motion \cite{mishra2019sub} assumption of the target implies that $\zeta_{m\rr \target n\rr }$ remains constant during each CPI. 

\subsubsection{IBFD MU-MIMO Communications}
\label{sec: comm_channels}
Denote the UL (DL) channel between the $\ith{i}$ UL UE ($\ith{j}$ DL UE) and BS as $\mathbf{H}_{i,\textrm{B}}\in\mathbb{C}^{\mathit{M}\cc\times N^{\textrm{u}}_{i}}$ ($\mathbf{H}_{\textrm{B},j}\in\mathbb{C}^{N^{\textrm{d}}_{j}\times \mathit{M}\cc}$), which is assumed to be full rank for all $i$ ($j$) to achieve the highest MIMO channel spatial DoF \cite{Multiuser}.  
The FD architecture implies that the BS Rx receives DL signals from the BS Tx through the \textit{self-interfering channel} $\mathbf{H}_{\mathrm{BB}}\in{\mathbb{C}^{N\cc\times \mathit{M}\cc}}$. The signals simultaneously transmitted by $\ith{i}$ UL UE interfere with the $j$-th DL UE in the channel $\mathbf{H}_{i,j}\in\mathbb{C}^{N^{\textrm{d}}_{j}\times N^{\textrm{u}}_{i}}$. 
\subsubsection{Joint Radar-Communications}
\label{sec: JRC_channels}
The concurrent transmission of MIMO radar and IBFD MU-MIMO communications system with overlapping spectrum implies that the communications Rxs experience interference from the radar transmissions and vice versa. For 
narrowband signaling, the channel impulse responses (CIRs) for paths BS-to-$n\rr$-th radar Rx, $i$-th UL UE-to-$n\rr$-th radar Rx, $m\rr$-th radar Tx-to-BS, and $m\rr$-th radar Tx-to-$j$-th DL UE are, respectively, \cite{Lops2020uplink} 
\begin{flalign}
\mathbf{g}_{\mathrm{Bm},n\rr}\paren{t,\tau}&=\boldsymbol{\alpha}_{\mathrm{Bm},n\rr}e^{j2\pi f_{\mathrm{Bm},n\rr}t}\delta\paren{\tau-\tau_{\mathrm{Bm},n\rr}},\label{eq: BS_radar_direct_channel}\\
	\mathbf{g}_{i,n\rr}\paren{t,\tau}&=\boldsymbol{\alpha}_{i,n\rr}e^{j2\pi f_{i,n\rr}t}\delta\paren{\tau-\tau_{i,n\rr}},\\
	\mathbf{g}_{m\rr,\B}\paren{t,\tau}&=\mathbf{h}_{m\rr, \B}e^{j2\pi f_{m\rr,\B}t}\delta\paren{\tau-\tau_{m\rr,\textrm{B}}},	\label{eq: radar_BS_channel}
\\
\textrm{and }\mathbf{g}_{m\rr,j}\paren{t,\tau}&=\mathbf{h}_{m\rr, j}e^{j2\pi f_{m\rr,j}t}\delta\paren{\tau-\tau_{m\rr,j}},\label{eq: radar_DL_channel}
\end{flalign}
where $\tau_{\mathrm{Bm},n\rr}$, $\tau_{i,n\rr}$, $\tau_{m\rr,\textrm{B}}$, and $\tau_{m\rr,j}$ denote the corresponding path delays; $f_{\textrm{Bm},n\rr}$, $f_{i,n\rr}$, $f_{m\rr,\B}$ and $f_{m\rr,j}$ are the respective Doppler shifts; $\boldsymbol{\alpha}_{\textrm{Bm},n\rr}\in\mathbb{C}^{M\cc}$ i.i.d. for $n\rr\in\mathbb{Z}_+\braces{N\rr}$, $\boldsymbol{\alpha}_{i,n\rr}\in\mathbb{C}^{N^{\textrm{u}}_i}$ i.i.d. for $n\rr\in\mathbb{Z}_+\braces{N\rr}$, $\mathbf{h}_{m\rr,\textrm{B}}\in\mathbb{C}^{N_{\mathrm{c}}}$ i.i.d. for $m\rr\in\mathbb{Z}_+\braces{M\rr}$, and $\mathbf{h}_{m\rr, j}\in\mathbb{C}^{N^{\textrm{d}}_j}$ i.i.d for $m\rr\in\mathbb{Z}_+\braces{M\rr}$ are complex channel vectors.

The proposed radar-communications co-design model exploits the DL signals to aid the radar target detection by enabling a cooperative DL-radar mode. \color{black}The moving target remains in the coverage area of the BS during the 
CPI. The CIR for the path BS-to-target-to-$n\rr$-th radar Rx  is
\begin{flalign}
\mathbf{h}_{\mathrm{Bt},n\rr}\paren{t,\tau}&=\alpha_{\mathrm{B\target},n\rr}\mathbf{a}_{\mathrm{T}}^\dagger\paren{\theta_{\mathrm{B}\target}}e^{j2\pi f_{\mathrm{B\target},n\rr}t}\delta\paren{\tau-\tau_{\mathrm{B\target},n\rr}},
\end{flalign}
where $\theta_{\mathrm{B}\target}$ is the AoD observed by the BS w.r.t the path BS-to-target-to-$n\rr$-th radar Rx;  $\mathbf{a}_{\mathrm{T}}^\dagger\paren{\theta_{\mathrm{B}\target}}$ is the corresponding transmit steering vector; 
and $\alpha_{\mathrm{Bt},n\rr}$, $\tau_{\mathrm{Bt},n\rr}$, and  $f_{\mathrm{Bt},n\rr}$ are the channel path gain, delay, and Doppler shift of the path BS-to-target-to-$n\rr$-th radar Rx, respectively. 

\section{Statistical MIMO Radar Receiver}
\label{sec:radar_total_receive}
The radar signals reflected off the targets and received at the radar Rxs are overlaid with clutter echoes, IBFD MU-MIMO signals, and noise. To retrieve the target reflected radar and communications signals, each radar Rx is equipped with $M\rr$ radar-specific matched filters $\braces{\phi^\ast_{m\rr}\bracket{-n},\forall m\rr}$ (matched to the waveforms $\braces{\phi_{m_\mathrm{r}}\bracket{n}, \forall m\rr}$, respectively) that operate in parallel with a  $G$-shifted receive pulse-shaping filter $p_{\textrm{R}}\bracket{n-G}$ 
(matched to $p_{\textrm{T}}\bracket{n}\triangleq p_{\textrm{T}}\paren{nT_\textrm{p}}$) to eliminate interference arising from simultaneous DL and UL communications. In practice, the discrete-time interval $G$ is estimated at both radar and communications Rxs through symbol synchronization methods based on, for instance, the maximum-likelihood estimator \cite{Multiuser}.

\subsection{Radar Signals Received at the Radar Rx}\label{sec: radar receive only} 
The target echo signals are downconverted at radar Rxs. Using the radar pulse train in \eqref{eq: radar_pulse_train} and $\alpha_{m\rr \target n\rr}$ defined in Section \ref{subsubsec:radar_channel}, the resulting baseband signal at the $\ith{n\rr}$ Rx in the observation window is 
\begin{flalign}
    \label{target return_CT}
    y_{\mathrm{rt},n\rr} \paren{t}&=\sum_{m_\mathrm{r}=1}^{M_\mathrm{r}}\alpha_{m_\mathrm{r}\target n_\mathrm{r}}s_{m_\mathrm{r}}(t-\zeta_{m\rr \target n\rr})\nonumber\\
    &=\sum_{m_\mathrm{r}=1}^{M_\mathrm{r}}\sum_{k=0}^{K-1} \alpha_{m_\mathrm{r}\target n_\mathrm{r}}a_{m_\mathrm{r},k}\phi_{m_\mathrm{r}}\paren{t-k T\rr-\zeta_{m\rr \target n\rr}-GT_{\mathrm{p}}}e^{j2\pi tf_{m_\mathrm{r}\target n_\mathrm{r}}}\nonumber\\
    &\approx\sum_{m_\mathrm{r}=1}^{M_\mathrm{r}}\sum_{k=0}^{K-1} \alpha_{m_\mathrm{r}\target n_\mathrm{r}}a_{m_\mathrm{r},k}\phi_{m_\mathrm{r}}\paren{t-k T\rr-\zeta_{m\rr \target n\rr}-GT_{\mathrm{p}}} e^{j2\pi k T_{\mathrm{r}}f_{m_\mathrm{r}\target n_\mathrm{r}}},
\end{flalign}
\normalsize
\color{black}
where the approximation follows the assumption $f_{m\rr\target n\rr}\ll\sfrac{1}{T_{\mathrm{p}}}$ so that the phase rotation is constant within one CPI \cite{hongbin_movingtarget,duggal2020doppler}.
Collect the exponential terms in a vector
\begin{align}
\mathbf{q}_{\mathrm{r},n\rr}\bracket{k}=\bracket{e^{j2\pi kT\rr f_{1 \target n\rr}},\cdots,e^{j2\pi kT\rr f_{M\rr \target n\rr}}}^\top\in\mathbb{C}^{M\rr},
\end{align}
and define 
\begin{align}
\mathbf{Q}\rnr\bracket{k}=\diag\paren{\mathbf{q}_{\mathrm{r},n\rr}\bracket{k}}. 
\end{align}
Sampling $y_{\mathrm{rt},n\rr}\paren{t}$ at the rate $\frac{1}{T_\mathrm{p}}$ yields
\begin{align}
y_{\mathrm{rt},n\rr}\bracket{n}=y_{\mathrm{rt},n\rr}\paren{nT_{\textrm{p}}}. 
\end{align}
Denote $d$ as the fast time index $d\in\mathbb{Z}\paren{N-1}$ and express the discrete-time version of \eqref{target return_CT} in terms of the slow and fast time indices as 
\begin{flalign}
y_{\mathrm{rt},n\rr}\bracket{k,d}&\triangleq y_{\mathrm{rt},n\rr}\bracket{n}|_{n=G+kN+d}\nonumber\\
&= y_{\mathrm{rt},n\rr}\bracket{G+kN+d}\nonumber\\
&=\sum_{m_\mathrm{r}=1}^{M_\mathrm{r}}e^{j2\pi kT_{\mathrm{r}}f_{m_\mathrm{r}\target n_\mathrm{r}}} \alpha_{m_\mathrm{r}\target n_\mathrm{r}}a_{m\rr,k}\phi_{m\rr}\paren{dT_\mathrm{p}-\zeta_{m\rr\target n\rr}}. 
\end{flalign}

Discretizing $\zeta_{m\rr\target n\rr}$ and following the narrowband assumption, we have the discrete-time delay index $n_{m\rr\target n\rr}=\left\lfloor\frac{\zeta_{m\rr\target n\rr}}{T_{\mathrm{p}}}\right\rfloor=n_\target$ for all $m\rr$ and $n\rr$\footnote{In general, the statistical MIMO radar receiver employs data association algorithms to ascertain the location and Doppler frequencies of each target using echoes from all Tx-Rx pairs. These algorithms are beyond the scope of this paper; we refer the reader to standard references, e.g., \cite{Nayebi13dataassociation}.}. Then, for the $\ith{k}$ PRI, the composite output of the $M\rr$ radar-specific matched filters at the $n_\target$-th range cell, which is referred to as the cell under test (CUT), of $n\rr$-th radar Rx is 
\begin{align}
y^{\paren{n_\target}}_{\mathrm{r},n\rr}\bracket{k}&\triangleq\sum_{m\rr=1}^{M\rr}y_{\mathrm{rt},n\rr}\bracket{k,d}\ast\phi_{m\rr}^\dagger\bracket{-d}|_{d=n_{\target}}\nonumber\\
&=\sum_{m_\mathrm{r}=1}^{M_\mathrm{r}}\sum_{m^\prime_\mathrm{r}=1}^{M_\mathrm{r}}\sum_{d^\prime=-\infty}^{\infty}e^{j2\pi kT_{\mathrm{r}}f_{m^\prime\rr\target n\rr}} \alpha_{m^\prime_\mathrm{r}\target n_\mathrm{r}}  a_{m^\prime\rr,k}\phi_{m_\mathrm{r}}\bracket{d^\prime-n_{ 
\target}}\phi^\ast_{m\rr}\bracket{d^\prime-d}\nonumber\\
&=\sum_{m_\mathrm{r}=1}^{M_\mathrm{r}}e^{j2\pi kT_{\mathrm{r}}f_{m_\mathrm{r}\target n_\mathrm{r}}} \alpha_{m_\mathrm{r}\target n_\mathrm{r}}a_{m\rr,k}=\mathbf{h}^\top_{\mathrm{rt},n\rr}\bracket{k}\mathbf{a}\bracket{k},
\end{align}
\normalsize
where  
\begin{align}
\mathbf{h}_{\mathrm{rt},n\rr}\bracket{k}=\bracket{h_{1\target,n\rr}\bracket{k},\cdots,h_{M\rr\target,n\rr}\bracket{k}}^\top\in\mathbb{C}^{M\rr}, 
\end{align}
denotes the target response observed at the $n\rr$-th radar Rx with \cite{20MISTAP} 
\begin{align}
    h_{m\rr\target,n\rr}\bracket{k}=\alpha_{m_\mathrm{r}\target n_\mathrm{r}}e^{j2\pi \paren{k-1}T_{\mathrm{r}}f_{m_\mathrm{r},\target,n_\mathrm{r}}}.
\end{align}
\color{black}
The filter $p_{\textrm{R}}\bracket{n}$ operates in parallel with $\braces{\phi^\ast_{m\rr}\bracket{-n}}$. Hence, $y_{\mathrm{rt},n\rr}\bracket{k,q}$ is also processed by $p_{\textrm{R}}\bracket{n-G}$, 
\begin{flalign}
v^{\paren{n_\target}}_{\textrm{r},n\rr}\bracket{k}&=y_{\mathrm{rt},n\rr}\bracket{k,q}\ast p_{\textrm{R}}\bracket{q-G}|_{q=n_{\target}}\nonumber\\
&=\sum_{m_\mathrm{r}=1}^{M_\mathrm{r}}e^{j2\pi \paren{k-1}T_{\mathrm{r}}f_{m_\mathrm{r}\target n_\mathrm{r}}} \alpha_{m_\mathrm{r}\target n_\mathrm{r}}a_{m\rr,k}\varphi_{m\rr,\textrm{R}}\bracket{G}\nonumber\\
&=\mathbf{h}^\top_{\textrm{rt},n\rr}\bracket{k}\paren{\boldsymbol{\varphi_{\textrm{R}}}\odot\mathbf{a}\bracket{k}},
\label{eq: radar_comm_match}
\end{flalign}\normalsize
where $\varphi_{m\rr,\textrm{R}}\bracket{n}$ is the cross-correlation function of $\phi_{m\rr}\bracket{n}$ and $p_{\textrm{R}}\bracket{n}$ and 
\begin{align}
\boldsymbol{\varphi_{\textrm{R}}}=\bracket{\varphi_{1,\textrm{R}}\bracket{G},\cdots,\varphi_{M\rr,\textrm{R}}\bracket{G}}\in\mathbb{R}^{M\rr}.
\end{align}
In the conventional MIMO Rx architecture, the target information is encapsulated in the signal $y^{\paren{n_\target}}_{\mathrm{r},n\rr}\bracket{k}$. However, $v^{\paren{n_\target}}_{\mathrm{r},n\rr}\bracket{k}$ also has useful target information and, if needed, may be used to increase the signal power. In that case, combining $y^{\paren{n_\target}}_{\mathrm{r},n\rr}\bracket{k}$ and $v^{\paren{n_\target}}_{\textrm{r},n\rr}\bracket{k}$ produces
\begin{align}
w^{\paren{n_\target}}_{\mathrm{r},n\rr}\bracket{k}=y^{\paren{n_\target}}_{\mathrm{r},n\rr}\bracket{k}+v^{\paren{n_\target}}_{\mathrm{r},n\rr}\bracket{k}=\mathbf{h}^\top_{\textrm{rt},n\rr}\bracket{k}\bracket{\paren{\boldsymbol{\varphi_{\textrm{R}}}+\mathbf{1}_{M\rr}}\odot\mathbf{a}\bracket{k}}.
\end{align}
With $G$ estimated, $\boldsymbol{\varphi_{\textrm{R}}}$ becomes a constant coefficient vector. Hence, without loss of generality, we derive the receive signal model and the CWSM algorithm based on  $y^{\paren{n_\target}}_{\textrm{r},n\rr}\bracket{k}$ without explicitly considering $\boldsymbol{\varphi_{\textrm{R}}}$. Stacking the samples of a CPI yields
\begin{equation}
\label{eq: radar_range_cell_total}
\mathbf{y}^{\paren{n_\target}}_{\mathrm{rt},n\rr}=\bracket{y^{\paren{n_\target}}_{\mathrm{rt},n\rr}\bracket{0},\cdots,y^{\paren{n_\target}}_{\mathrm{rt},n\rr}\bracket{K-1}}^\top\in\mathbb{C}^{K}.    
\end{equation}  \normalsize
\color{black}

Denote the covariance matrix (CM) of $\mathbf{y}_{\mathrm{rt},n\rr}$ by $\mathbf{R}_{\mathrm{rt},n\rr}\in\mathbb{C}^{K\times K}$, whose $\ith{\paren{m,l}}$ element is
\begin{align}
\mathbf{R}_{\mathrm{rt},n\rr}\paren{m,l}&=\mathbb{E}\bracket{y^{\paren{n_\target}}_{\mathrm{r},n\rr}\bracket{m}\paren{y^{\paren{n_\target}}_{\mathrm{r},n\rr}\bracket{l}}^\dagger}=\trace\braces{\mathbf{a}\bracket{m}\mathbf{a}^\dagger\bracket{l}\boldsymbol{\Sigma}^{\paren{m,l}}_{\mathrm{rt},n\rr}},
\end{align}
where $\boldsymbol{\Sigma}^{\paren{m,l}}_{\mathrm{r}n_\target,n\rr}=\mathbb{E}\braces{\mathbf{h}^\ast_{\mathrm{rt},n\rr}\bracket{l}\mathbf{h}^\top_{\mathrm{rt},n\rr}\bracket{m}}$ is a diagonal matrix with $\ith{m\rr}$ diagonal element as $e^{j2\pi T_{\textrm{r}}\paren{m-l}f_{m\rr\target n\rr}}\eta^2_{m\rr\target n\rr}$.

\subsection{Communications Signals at the Radar Rx}
\label{sec:co-existence}
Using CIRs $\mathbf{g}_{\mathrm{Bm},n\rr}$ and $\mathbf{h}_{\mathrm{Bt},n\rr}$, the DL signals at $n\rr$-th radar Rx are 
\begin{flalign}
\widetilde{y}_{\mathrm{Bm},n\rr}\paren{t}&=\int_{\tau}\mathbf{g}_{\mathrm{Bm},n\rr}^T\paren{t,\tau}\mathbf{x}_{\mathrm{B}}\paren{t-\tau}d\tau\nonumber\\
&=\boldsymbol{\alpha}^\top_{\mathrm{Bm},n\rr}e^{j2\pi f_{\mathrm{Bm},n\rr}t}\mathbf{x}_{\mathrm{B}}\paren{t-\tau_{\mathrm{Bm},n\rr}}, 
\end{flalign}
and
\begin{flalign}
\widetilde{y}_{\mathrm{Bt},n\rr}\paren{t}&=\int_{\tau}\mathbf{h}_{\mathrm{Bt},n\rr}^T\paren{t,\tau}\mathbf{x}_{\mathrm{B}}\paren{t-\tau}d\tau\nonumber\\
&=\alpha_{\mathrm{B\target},n\rr}\mathbf{a}_{\mathrm{T}}^\dagger\paren{\theta_{\mathrm{B}\target}}e^{j2\pi f_{\mathrm{B\target},n\rr}t}\mathbf{x}_{\mathrm{B}}\paren{t-\tau_{\mathrm{Bt},n\rr}}, 
\end{flalign}
\normalsize
respectively. 
The $i$-th UL UE signal received by $n\rr$-th Rx is  
\begin{flalign}
\widetilde{y}_{i,n\rr}\paren{t}&=\int_{\tau}\mathbf{g}_{i,n\rr}\paren{t,\tau}\mathbf{x}_{\mathrm{u},i}\paren{t-\tau}d\tau\nonumber\\
&=\boldsymbol{\alpha}^\top_{i,n\rr}e^{j2\pi f_{i,n\rr}t}\mathbf{x}_{\mathrm{u},i}\paren{t-\tau_{i,n\rr}}.
\end{flalign}
Discretizing $\tau_{\mathrm{Bm},n\rr}$, $\tau_{\mathrm{Bt},n\rr}$, and $\tau_{i,n\rr}$, and invoking narrowband assumption, we have $\lfloor\sfrac{\tau_{\mathrm{Bm},n\rr}}{T_\mathrm{p}}\rfloor=n_{\mathrm{Bm}}$, $\lfloor\sfrac{\tau_{\mathrm{Bt},n\rr}}{T_\mathrm{p}}\rfloor=n_{\mathrm{t}}$, $\lfloor\sfrac{\tau_{i,n\rr}}{T_\mathrm{p}}\rfloor=n_{\mathrm{u}}$ for all $i$ and $n\rr$. Sampling $\widetilde{y}_{\mathrm{Bm},n\rr}\paren{t}$, $\widetilde{y}_{\mathrm{Bt},n\rr}\paren{t}$, and $\widetilde{y}_{i,n\rr}\paren{t}$ at $\sfrac{1}{T_{\mathrm{p}}}$ produces $\widetilde{y}_{\mathrm{Bm},n\rr}\bracket{n}=\widetilde{y}_{\mathrm{Bm},n\rr}\paren{nT_{\mathrm{p}}}$, $\widetilde{y}_{\mathrm{Bt},n\rr}\bracket{n}=\widetilde{y}_{\mathrm{Bt},n\rr}\paren{nT_{\mathrm{p}}}$ and $\widetilde{y}_{i,n\rr}\bracket{n}=\widetilde{y}_{i,n\rr}\paren{nT_{\mathrm{p}}}$. 
As a result, the output of $p_{\mathrm{R}}\bracket{n}$ w.r.t. $\widetilde{y}_{\mathrm{Bm},n\rr}\bracket{n}$, $\widetilde{y}_{\mathrm{Bt},n\rr}\bracket{n}$ and $\widetilde{y}_{i,n\rr}\bracket{n}$ are
\begin{flalign}
y_{\mathrm{Bm},n\rr}\bracket{n}&=\widetilde{y}_{\mathrm{Bm},n\rr}\bracket{n}\ast p_{\mathrm{R}}\bracket{n-G}\nonumber\\
&=\boldsymbol{\alpha}^\top_{\textrm{Bm},n\rr}\sum_{k=0}^{K-1}e^{j2\pi f_{\mathrm{Bm},n\rr}kNT_{\mathrm{p}}}\sum_{l=0}^{N-1}\mathbf{s}_{\mathrm{B}}\bracket{k,l} p_{\mathrm{TR}}\bracket{n-G-n_{\mathrm{Bm}}-\paren{k N+l}},\label{eq: Bm_nr_1}\\
y_{\mathrm{Bt},n\rr}\bracket{n}
&=\alpha_{\mathrm{Bt},n\rr}\mathbf{a}_{\mathrm{T}}^\dagger\paren{\theta_{\mathrm{Bt}}}\sum_{k=0}^{K-1}e^{j2\pi f_{\mathrm{Bt},n\rr}kNT_{\mathrm{p}}}\sum_{l=0}^{N-1}\mathbf{s}_{\mathrm{B}}\bracket{k,l} p_{\mathrm{TR}}\bracket{n-G-n_{\mathrm{Bt}}-\paren{k N+l}},\label{eq: Bt_nr_1}\\
\textrm{and }
y_{i,n\rr}\bracket{n}
&=\boldsymbol{\alpha}^\top_{i,n\rr}\sum_{k=0}^{K-1}e^{j2\pi f_{i,n\rr}kT_\mathrm{r}}\sum_{l=0}^{N-1}\mathbf{s}_{\mathrm{u},i}\bracket{k,l} p_{\mathrm{TR}}\bracket{n-G-n_{\mathrm{u}}-\paren{k N+l}},\label{eq: UL_nr_1}
\end{flalign}\normalsize
respectively, where $p_{\mathrm{TR}}\bracket{n}=p_{\mathrm{T}}\bracket{n}\ast p_\mathrm{R}\bracket{n}$ denotes the transmit-receive pulse shaping filter satisfying the Nyquist criterion \cite{Multiuser,duggal2020doppler}. Therefore, the optimal sampling times associated with \eqref{eq: Bm_nr_1}, \eqref{eq: Bt_nr_1} and \eqref{eq: UL_nr_1} that produce zero ISI and yield $p_{\mathrm{TR}}\bracket{n}\neq0$  are  
$n = kN+l+G+n_{\mathrm{Bm}}$, $n=kN+l+G+n_{\mathrm{t}}$, and $n=kN+l+G+n_{\mathrm{u}}$, respectively. In practice, the root-raised-cosine filter is commonly used for $p_\textrm{T}\bracket{n}$ and $p_{\textrm{R}}\bracket{n}$; their product is the raised cosine function, which is a Nyquist filter \cite{Multiuser}. 

We are interested in the communications symbols appearing in the CUT of radar Rxs, i.e., $n=kN+G+n_\target$, which leads to $l=0$, in \eqref{eq: Bt_nr_1}. Note that standards such as IEEE 802.11ad [45] 
have utilized training symbols (sent at the beginning of each frame) of DL signals for radar sensing functions. For our co-design paradigm,  $\mathbf{s}_{\textrm{B}}\bracket{k,0}$ is a training symbol vector known to the radar Rxs for all $k$ for our co-design paradigm. Without loss of generality, we assume that the target is located further than either the BS or UL UEs from the radar Rxs, i.e., $n_{\mathrm{Bm}}<n_{\mathrm{t}}<N$ and $n_{\mathrm{u}}<n_{\mathrm{t}}<N$. This guarantees that only the $\ith{k}$ communications frames arrive in the $\ith{k}$ radar PRI. The DL and UL symbol indices corresponding to $y_{\mathrm{Bm},n\rr}\bracket{n}$ and $y_{i,n\rr}\bracket{n}$ observed in the CUT of the $\ith{n\rr}$ radar Rx in the $\ith{k}$ PRI  
are $l=n_\target-n_{\mathrm{Bm}}$ in \eqref{eq: Bm_nr_1} and $l=n_\target-n_{\mathrm{u}}$ in \eqref{eq: UL_nr_1}. Therefore, the $\ith{n\rr}$ radar Rx recovers the communications symbol vectors $\mathbf{s}^{\paren{n_\target}}_{\mathrm{Bm},n\rr}\bracket{k}=   \mathbf{s}_{\mathrm{B}}\bracket{k,\paren{n_\target-n_{\mathrm{Bm}}}}$, $\mathbf{s}_{\mathrm{Bt},n\rr}\bracket{k}=\mathbf{s}_{\mathrm{B}}\bracket{k,0}$, and $\mathbf{s}^{\paren{n_\target}}_{i,n\rr}\bracket{k}=   \mathbf{s}_{\mathrm{u},i}\bracket{k,\paren{n_\target-n_{\mathrm{u}}}}$ for all $i$ in its CUT during the $\ith{k}$ PRI. 
Hence, the communications signal components, i.e. DL, target echo via DL, and UL, at range cell $n_\target$ of the $\ith{n\rr}$ radar Rx in the $\ith{k}$ PRI are, respectively, \footnote{\textcolor{black}{In practice, the symbol indices $n_{\textrm{Bm}}$ and $n_{\textrm{u}}$ are communicated to radar Rxs via either a fusion center \cite{MCMIMO_RadComm,rihan2018optimum} or direct feedback from the BS \cite{interferencealignment}.}} 
\begin{flalign}
y^{\paren{n_\target}}_{\mathrm{Bm},n\rr}\bracket{k}&\triangleq y_{\mathrm{Bm},n\rr}\bracket{kN+n_\target} =\boldsymbol{\alpha}^\top_{\textrm{Bm},n\rr}e^{j2\pi f_{\mathrm{Bm},n\rr}kT_{\mathrm{r}}}\mathbf{s}^{\paren{n_\target}}_{\mathrm{Bm},n\rr}\bracket{k}\nonumber\\
&=\mathbf{h}^\top_{\mathrm{Bm},n\rr}\bracket{k}\mathbf{s}^{\paren{n_\target}}_{\mathrm{Bm},n\rr}\bracket{k},\nonumber\\
y^{\paren{n_\target}}_{\mathrm{Bt},n\rr}\bracket{k}
&=\mathbf{h}^\top_{\mathrm{Bt},n\rr}\bracket{k}\mathbf{s}^{\paren{n_\target}}_{\mathrm{Bt},n\rr}\bracket{k},\\
\textrm{and } 
y^{\paren{n_\target}}_{\mathrm{u},n\rr}\bracket{k}
&=\sum_{i=1}^{I}\mathbf{h}^\top_{i,n\rr}\bracket{k}\mathbf{s}^{\paren{n_\target}}_{i,n\rr}\bracket{k},
\end{flalign}
\normalsize 
where $\mathbf{h}_{\mathrm{Bt},n\rr}\bracket{k}$ is the DL target response and $\mathbf{h}_{\mathrm{Bm},n\rr}\bracket{k}$ and $\mathbf{h}_{i,n\rr}\bracket{k}$ are the DL and direct path responses, respectively.  

Following the discussion after \eqref{eq: radar_comm_match} in Section~\ref{sec: radar receive only}, we omit the communications signal components at the $\phi^\ast_{m\rr}\bracket{-n}$ outputs at the radar Rxs. Stacking samples 
from a CPI in  vectors gives
\begin{align}
\label{eq: Bm_range_cell}
\mathbf{y}^{\paren{n_\target}}_{\mathrm{Bm},n\rr}&=\bracket{y^{\paren{n_\target}}_{\mathrm{Bm},n\rr}\bracket{0},\cdots,y^{\paren{n_\target}}_{\mathrm{Bm},n\rr}\bracket{k},\cdots,y^{\paren{n_\target}}_{\mathrm{Bm},n\rr}\bracket{K-1}}^\top,\\
\mathbf{y}^{\paren{n_\target}}_{\mathrm{Bt},n\rr}&=\bracket{y^{\paren{n_\target}}_{\mathrm{Bt},n\rr}\bracket{0},\cdots,y^{\paren{n_\target}}_{\mathrm{Bt},n\rr}\bracket{k},\cdots,y^{\paren{n_\target}}_{\mathrm{Bt},n\rr}\bracket{K-1}}^\top, \label{eq:Bt_range_cell}\\
\textrm{and }
\mathbf{y}^{\paren{n_\target}}_{\mathrm{u},n\rr}&=\bracket{y^{\paren{n_\target}}_{\mathrm{u},n\rr}\bracket{0},\cdots,y^{\paren{n\target}}_{\mathrm{u},n\rr}\bracket{k},\cdots,y^{\paren{n_\target}}_{\mathrm{u},n\rr}\bracket{K-1}},\label{eq:Ur_range_cell}
\end{align}

Denote the CMs of $\mathbf{y}^{\paren{n_\target}}_{\mathrm{Bm},n\rr}$, $\mathbf{y}^{\paren{n_\target}}_{\mathrm{Bt},n\rr}$, and $\mathbf{y}^{\paren{n_\target}}_{\mathrm{u},n\rr}$ by $\mathbf{R}^{\paren{n_\target}}_{\mathrm{Bm},n\rr}$, $\mathbf{R}^{\paren{n_\target}}_{\mathrm{Bt},n\rr}$, $\mathbf{R}^{\paren{n_\target}}_{\textrm{UL},n\rr}$, respectively, whose $\ith{\paren{m,\ell}}$ elements are 
\begin{flalign}
\mathbf{R}^{\paren{n_\target}}_{\mathrm{Bm},n\rr}\paren{m,\ell}&
=\trace\braces{\mathbf{s}^{\paren{n_\target}}_{\mathrm{Bm},n\rr}\bracket{m}\paren{\mathbf{s}^{\paren{n_\target}}_{\mathrm{Bm},n\rr}\bracket{\ell}}^\dagger\boldsymbol{\Sigma}^{\paren{m,\ell}}_{\mathrm{Bm},n\rr}},\\
\mathbf{R}^{\paren{n_\target}}_{\mathrm{Bt},n\rr}\paren{m,\ell}
&=\trace\braces{\mathbf{s}^{\paren{n_\target}}_{\mathrm{Bt},n\rr}\bracket{m}\paren{\mathbf{s}^{\paren{n_\target}}_{\mathrm{Bt},n\rr}\bracket{\ell}}^\dagger\boldsymbol{\Sigma}^{\paren{m,\ell}}_{\mathrm{Bt},n\rr}},\\
\text{and\;
}\mathbf{R}^{\paren{n_\target}}_{\textrm{UL},n\rr}\paren{m,\ell}
&=\sum_{i=1}^I\trace\braces{\mathbf{s}^{\paren{n_\target}}_{i,n\rr}\bracket{m}\paren{\mathbf{s}^{\paren{n_\target}}_{i,n\rr}\bracket{\ell}}^\dagger\boldsymbol{\Sigma}^{\paren{m,\ell}}_{i,n\rr}},
\end{flalign}
\normalsize
where 
\begin{align}
\boldsymbol{\Sigma}^{\paren{m,\ell}}_{\mathrm{Bm},n\rr}&=\mathbb{E}\bracket{\mathbf{h}^\ast_{\mathrm{Bm},n\rr}\bracket{\ell}\mathbf{h}^\top_{\mathrm{Bm},n\rr}\bracket{m}},\nonumber\\
\boldsymbol{\Sigma}^{\paren{m,\ell}}_{\mathrm{Bt},n\rr}&=\mathbb{E}\bracket{\mathbf{h}^\ast_{\mathrm{Bt},n\rr}\bracket{\ell}\mathbf{h}^\top_{\mathrm{Bt},n\rr}\bracket{m}},\nonumber\\ \textrm{and } \boldsymbol{\Sigma}^{\paren{m,\ell}}_{i,n\rr}&=\mathbb{E}\bracket{\mathbf{h}^\ast_{i,n\rr}\bracket{\ell}\mathbf{h}^\top_{i,n\rr}\bracket{m}}.
\end{align}
Here, an average is taken over the channels because, for the radar received signal, the target information is the source of the randomness. 

\subsection{Combined target return from radar and DL reflected signals}
In order to enhance the radar target detection, we utilize \eqref{eq:Bt_range_cell} in processing it jointly with \eqref{eq: radar_range_cell_total}. 
Denote the combined target reflected signal received at the $n\rr$-th radar Rx  in the $\ith{k}$ PRI as 
\begin{flalign}
\label{eq:target1}
y^{\paren{n_\target}}_{\mathrm{t},n\rr}\bracket{k}&=y^{\paren{n_\target}}_{\mathrm{rt},n\rr}\bracket{k}+y^{\paren{n_\target}}_{\mathrm{Bt},n\rr}\bracket{k}=\mathbf{h}^\top_{\mathrm{rt},n\rr}\bracket{k}\mathbf{a}\bracket{k}+\mathbf{h}^\top_{\mathrm{Bt},n\rr}\bracket{k}\mathbf{s}^{\paren{n_\target}}_{\mathrm{Bt},n\rr}\bracket{k}\nonumber\\
&=\mathbf{h}^\top_{\target,n\rr}\bracket{k}\mathbf{s}^{\paren{n_\target}}_{\mathrm{t},n\rr}\bracket{k},
\end{flalign}\normalsize
where $\mathit{M}=\mathit{M}\cc+\MM\rr$ is the number of effective transmit antennas for target detection,
\begin{align}
\mathbf{h}_{\mathrm{t},n\rr}\bracket{k}=\bracket{\mathbf{h}^\top_{\mathrm{rt},n\rr}\bracket{k},\mathbf{h}^\top_{\mathrm{Bt},n\rr}\bracket{k}}^\top=\mathbf{J}\rr\mathbf{h}_{\mathrm{rt},n\rr}\bracket{k}+\mathbf{J}_\B\mathbf{h}_{\mathrm{Bt},n\rr}\bracket{k},
\end{align}
denotes the total target response observed at the $n\rr$-th radar Rx and
\begin{align}\mathbf{s}^{\paren{n_\target}}_{\mathrm{t},n\rr}\bracket{k}=\bracket{\mathbf{a}^\top\bracket{k},\paren{\mathbf{s}^{\paren{n_\target}}_{\mathrm{Bt},n\rr}\bracket{k}}^\top}^\top=\mathbf{J}\rr\mathbf{a}\bracket{k}+\mathbf{J}_\B\mathbf{s}^{\paren{n_\target}}_{\mathrm{Bt},n\rr}\bracket{k}\end{align} is the effective transmit signal vector, $\mathbf{J}_{\textrm{r}}=\bracket{\mathbf{I}_{\mathit{M}\rr\times \mathit{M}\rr};\mathbf{0}_{\mathit{M}\cc\times \mathit{M}\rr}}\in\mathbb{R}^{\mathbf{\mathit{M}\times \mathit{M}\rr}}$ and $\mathbf{J}_{\textrm{B}}=\bracket{\mathbf{0}_{\mathit{M}\rr\times \mathit{M}\cc};\mathbf{I}_{\mathit{M}\cc\times \mathit{M}\cc}}\in\mathbb{R}^{\mathbf{\mathit{M}\times \mathit{M}\cc}}$. Hence the $\ith{\paren{k,l}}$ element of $\mathbf{R}_{\target,n\rr}\in\mathbb{C}^{K\times K}$ is 
\begin{flalign}
\mathbf{R}^{\paren{n_\target}}_{\target,n\rr}\paren{k,l}=\mathbf{R}^{\paren{n_\target}}_{\mathrm{rt},n\rr}\paren{k,l}+\mathbf{R}^{\paren{n_\target}}_{\mathrm{Bt},n\rr}\paren{k,l}.
\end{flalign}\normalsize

As the MIMO radar code matrix is based on the output of matched filtering at the CUT, we drop the cell index $n_\target$ in the sequel for simplicity. For the entire CPI, we have
\begin{flalign}
\mathbf{y}_{\mathrm{t},n\rr}&=\bracket{y_{\mathrm{t},n\rr}\bracket{0},\cdots,y_{\mathrm{t},n\rr}\bracket{K-1}}^\top=\mathbf{S}_{\mathrm{t},n\rr}\mathbf{h}_{\mathrm{t},n\rr}
\end{flalign}\normalsize
where  
\begin{align}
\mathbf{S}_{\mathrm{t},n\rr}&=\oplus_{k=1}^{K}\mathbf{s}^\top_{\target,n\rr}\bracket{k},\;\in\mathbb{C}^{K\times KM},\\ \mathbf{h}_{\mathrm{t},n\rr}&=\bracket{\mathbf{h}^\top_{\target,n\rr}\bracket{0},\cdots,\mathbf{h}^\top_{\target,n\rr}\bracket{K-1}}^\top \nonumber\\
&= \sum_{k=1}^{K}\mathbf{J}_{\textrm{h}}\bracket{k}\mathbf{h}_{\target,n\rr}\bracket{k},\;\in\mathbb{C}^{KM},
\end{align}
and 
\begin{align}
\mathbf{J}_{\mathrm{h}}\bracket{k}&=\bracket{\mathbf{0}_{\paren{k-1}M\times M};\mathbf{I}_{M};\mathbf{0}_{\paren{K-k}M\times M}},\;\in\mathbb{Z}^{KM\times M}.
\end{align}
Then, $\mathbf{R}_{\target,n\rr}=\mathbf{S}_{\target,n\rr}\boldsymbol{\Sigma}_{\target,n\rr}\mathbf{S}^\dagger_{\target,n\rr}$, where $\boldsymbol{\Sigma}_{\target,n\rr}=\mathbb{E}\bracket{\mathbf{h}_{\target,n\rr}\mathbf{h}^\dagger_{\target,n\rr}}\in\mathbb{C}^{KM\times KM}$.

\subsection{Clutter echoes}
In practice, apart from the target, the MIMO radar Rxs also receive echoes from undesired targets or clutter such as buildings and forests. The clutter echoes are treated as signal-dependent interference produced by many independent and unambiguous point-like scatterers \cite{NaghshTSP2017}. Denote the clutter trail at the CUT of $\ith{n\rr}$ Rx in the $\ith{k}$ PRI by 
\begin{align}
    y_{\textrm{c},n\rr}\bracket{k}=\sum_{m\rr=1}^{M\rr}\rho_{m\rr\textrm{c}n\rr}a_{m\rr,k}=\boldsymbol{\rho}^\top_{\textrm{r},n\rr}\mathbf{a}\bracket{k} 
\end{align}
where $\rho_{m\rr\textrm{c}n\rr}\sim\mathcal{CN}\paren{0,\sigma^2_{m\rr\textrm{c}n\rr}}$ denotes the clutter component reflection coefficient associated with the path between the $m\rr$-th radar Tx and $n\rr$-th radar Rx  and is the $\ith{m\rr}$ element of $\boldsymbol{\rho}_{\textrm{c},n\rr}\in\mathbb{C}^{M\rr}$. For a CPI, we have $\mathbf{y}_{\textrm{c},n\rr}=\mathbf{A}\boldsymbol{\rho}_{\textrm{c},n\rr}\in\mathbb{C}^{K}$, whose CM is obtained as
\begin{align}
\mathbf{R}_{\textrm{c},n\rr}=\mathbf{A}\boldsymbol{\Sigma}_{\textrm{c},n\rr}\mathbf{A}^\dagger,
\end{align}
with its $\ith{\paren{m,l}}$ element being
\begin{align}
\mathbf{R}_{\textrm{c},n\rr}\paren{m,l}=\trace\braces{\mathbf{a}\bracket{m}\mathbf{a}^\dagger\bracket{l}\boldsymbol{\Sigma}_{\textrm{c},n\rr}},
\end{align}
where $\boldsymbol{\Sigma}_{\textrm{c},n\rr}=\mathbb{E}\bracket{\boldsymbol{\rho}_{\textrm{c},n\rr}\boldsymbol{\rho}^\dagger_{\textrm{c},n\rr}}$.

\subsection{Composite radar received signal}
With the CSCG noise vector at the $\ith{n\rr}$ radar Rx by $\mathbf{z}\rnr\in\mathcal{CN}\paren{\mathbf{0},\sigma^2\rnr\mathbf{I}_{K}}$, the composite receive signal model at the CUT of the $\ith{n\rr}$ radar Rx is
\begin{flalign}
\mathbf{y}\rnr=\mathbf{y}_{\mathrm{t},n\rr}+\underbrace{\mathbf{y}_{\mathrm{c},n\rr}+\mathbf{y}_{\mathrm{Bm},n\rr}+\mathbf{y}_{\mathrm{u},n\rr}+\mathbf{z}\rnr}_{=\mathbf{y}^{\mathrm{in}}_{\mathrm{r},n\rr}},\label{eq:combined_rad_rx}
\end{flalign}\normalsize
where $\mathbf{y}^{\mathrm{in}}_{\mathrm{r},n\rr}$ denotes the interference-plus-noise component of $\mathbf{y}\rnr$. The variables $\mathbf{h}_{\mathrm{t},n\rr}$, $\boldsymbol{\alpha}_{\mathrm{Bm},n\rr}$, $\boldsymbol{\alpha}_{i,n\rr}$, $\boldsymbol{\rho}_{\mathrm{c},n\rr}$, and $\mathbf{z}\rnr$ are statistically independent; thus, the CM of $\mathbf{y}\rnr$ is 
\begin{align}
\mathbf{R}_{\textrm{r},n\rr}=\mathbf{R}_{\target,n\rr}+\mathbf{R}^{\mathrm{in}}_{\mathrm{r},n\rr},
\end{align}
with the CM of $\mathbf{y}^{\mathrm{in}}_{\textrm{r},n\rr}$ given by
\begin{align}
    \mathbf{R}^{\mathrm{in}}_{\mathrm{r},n\rr}\triangleq\mathbf{R}_{\textrm{c},n\rr}+\mathbf{R}_{\mathrm{Bm},n\rr}+\mathbf{R}_{\mathrm{Ur},n\rr}+\sigma^2\rnr\mathbf{I}_{K}.
\end{align} 

Combining the received signals from $N\rr$ radar Rxs yields
\begin{align}
\mathbf{y}_{\mathrm{r}}=\mathbf{y}_{\textrm{tr}}+\mathbf{y}^{\textrm{in}}_{\textrm{r}}=\bracket{\mathbf{y}^\top_{\textrm{r},1};\cdots;\mathbf{y}^\top_{\textrm{r},N\rr}}^\top\in\mathbb{C}^{KN\rr},
\end{align}
where 
\begin{align}
\mathbf{y}_{\textrm{tr}}=\bracket{\mathbf{y}^\top_{\textrm{t},1};\cdots;\mathbf{y}^\top_{\textrm{t},N\rr}}^\top,
\end{align}
and 
\begin{align}
\mathbf{y}^{\textrm{in}}_{\textrm{r}}\triangleq\mathbf{y}_{\textrm{cr}}+\mathbf{y}_{\textrm{Bmr}}+\mathbf{y}_{\textrm{Ur}}+\mathbf{z}_{\textrm{r}}=\bracket{\paren{\mathbf{y}^{\textrm{in}}_{\textrm{r},1}}^\top;\cdots;\paren{\mathbf{y}^{\textrm{in}}_{\textrm{r},N\rr}}^\top}^\top,
\end{align}
whose CM is 
\begin{align}
\mathbf{R}^{\textrm{in}}_{\textrm{r}}=\oplus_{n\rr=1}^{N\rr}\mathbf{R}^{\textrm{in}}_{\textrm{r},n\rr}.
\end{align}
\color{black}
\section{IBFD MU-MIMO Communications Receiver}
\label{sec:comm_rx}
Within the observation window, $J$ DL UEs and the BS receive both IBFD communications signals and radar probing signals. \color{black} The communications Rxs are equipped with receive filters matched with the $G$-shifted radar waveforms $\phi_{m\rr}\bracket{n+G}$ for all $m\rr$ operating in parallel with $p_{\textrm{R}}\bracket{n}$ to separate radar signals from communications. \color{black}  
\subsection{FD Communications Signals at Communications Rxs}\label{sec: comm_received}
The received signal 
at the BS Rx from the $\ith{i}$ UL UE is $\mathbf{y}_{i,\mathrm{B}}\paren{t}=\mathbf{H}_{i,\textrm{B}}\mathbf{x}_{\mathrm{u},i}\paren{t}$. Sampling $\mathbf{y}_{i,\mathrm{B}}\paren{t}$ at symbol rate  $\sfrac{1}{T_{\mathrm{p}}}$ yields 
\begin{align}
\mathbf{y}_{i,\mathrm{B}}\bracket{n}=\mathbf{H}_{i,\textrm{B}}\mathbf{x}_{\mathrm{u},i}\paren{nT_{\mathrm{p}}}\triangleq\mathbf{y}_{i,\mathrm{B}}\paren{nT_{\mathrm{p}}}.
\end{align}
When $\mathbf{y}_{i,\mathrm{B}}\bracket{n}$ is processed by $p_{\textrm{R}}\bracket{n}$, the output at $\ith{l}$ symbol period of the $\ith{k}$ frame is 
\begin{flalign}
\label{eq:ULFDcomm}
\mathbf{y}_{i,\mathrm{B}}\bracket{k,l}&\triangleq\mathbf{y}_{i,\mathrm{B}}\bracket{kN+l}\nonumber\\
&=\mathbf{H}_{i,\textrm{B}}\mathbf{x}_{\mathrm{u},i}\bracket{n}\ast p_{\textrm{R}}\bracket{n}|_{n=kN+l}\nonumber\\
&=\sum_{l^\prime=0}^{N-1}\mathbf{H}_{i,\textrm{B}}\mathbf{s}_{\mathrm{u},i}\bracket{k,l}p_{\mathrm{TR}}\bracket{kN+l-\paren{k N+l^\prime}}\nonumber\\
&=\mathbf{H}_{i,\textrm{B}}\mathbf{s}_{\textrm{u},i}\bracket{k,l}.
\end{flalign}
\normalsize
The simultaneous transmissions of all UL UEs lead to the MU interference (MUI)\footnote{The MU transmission here is via linear precoders, which have lower complexity than optimal dirty paper coding \cite{Multiuser}.}. Following $\paren{\ref{eq:ULFDcomm}}$ and assuming that the time/frequency synchronization is achieved, the MUI signal of the $\ith{i}$ UL UE is 
\begin{align}
\mathbf{y}_{\textrm{um},i}\bracket{k,l}=\sum_{q\neq i}\mathbf{H}_{q,\textrm{B}}\mathbf{s}_{\textrm{u},q}\bracket{k,l}.\label{eq:mui}
\end{align}\normalsize
Similarly, the DL signal observed at the BS Rx through the self-interfering channel is 
\begin{align}
\mathbf{y}_{\mathrm{BB}}\bracket{k,l}=\mathbf{H}_{\mathrm{BB}}\sum_{j=1}^{\mathit{J}}\PBj\mathbf{d}_{\mathrm{d},j}\bracket{k,l}. \label{eq:sic}
\end{align}

The CMs of 
$\mathbf{y}_{i,\textrm{B}}\bracket{k,l}$, $\mathbf{y}_{\textrm{um},i}\bracket{k,l}$,  and $\mathbf{y}_{\mathrm{BB}}\bracket{k,l}$ are, respectively,
\begin{align}
\mathbf{R}_{\textrm{u},i}\bracket{k,l}&=\HiB\PiB\PiBH\HiBH,\\
\mathbf{R}_{\textrm{um},i}\bracket{k,l}&=\sum_{g\neq i }\mathbf{H}_{g,\textrm{B}}\mathbf{P}_{\textrm{u},g}\mathbf{P}^\dagger_{\textrm{u},g}\mathbf{H}^\dagger_{g,\textrm{B}}, \\
\textrm{and }
\mathbf{R}_{\mathrm{BB}}\bracket{k,l}&=\sum_{j=1}^{\mathit{J}}\mathbf{H}_{\mathrm{BB}}\PBj\PBjH\mathbf{H}^\dagger_\mathrm{BB}.
\end{align}
Here, the source of the randomness lies in the symbol vector. Hence, the average is taken over the symbols.

Similar to $\paren{\ref{eq:ULFDcomm}}$, the discrete-time DL signal sampled at the $\ith{l}$ symbol period of the $\ith{k}$ frame by the  $\ith{j}$ DL UE is 
\begin{flalign}
\label{eq:DL1}
\mathbf{y}_{\textrm{B},j}\bracket{k,l}&=\mathbf{H}_{\textrm{B},j}\mathbf{x}_{\textrm{B}}\bracket{n}\ast p_\textrm{R}\bracket{n}|_{n=mN+l}
=\mathbf{H}_{\textrm{B},j}\mathbf{s}_{\textrm{d},j}\bracket{k,l}+\mathbf{y}_{\textrm{dm},j}\bracket{k,l},
\end{flalign}\normalsize
\color{black}where 
\begin{align}
\mathbf{y}_{\textrm{dm},j}\bracket{k,l}=\mathbf{H}_{\textrm{B},j}\sum_{g\neq j}^{}\mathbf{s}_{\textrm{d},g}\bracket{k,l}
\end{align}
denotes the MUI of the $\ith{j}$ DL UE. The UL interfering signal received at the $\ith{j}$ DL UE during the $\ith{l}$ symbol period of the $\ith{k}$ frame is 
\begin{align}
\mathbf{y}_{\mathrm{u},j}\bracket{k,l}=\sum_{i=1}^{\mathit{I}}\mathbf{H}_{i,j}\PiB\dui. \label{eq:UL2}
\end{align}\normalsize
The symbol vectors $\mathbf{d}_{\mathrm{d},j}$ are i.i.d. for all $j$. Hence, the CMs of $\mathbf{y}_{\textrm{B},j}\bracket{k,l}$, $\mathbf{y}_{\textrm{dm},j}\bracket{k,l}$, and $\mathbf{y}_{\mathrm{u},j}\bracket{k,l}$ are, respectively,
\begin{align}
\mathbf{R}_{\B,j}\bracket{k,l}&=\mathbf{H}_{\textrm{B},j}\PBj\PBjH\mathbf{H}^\dagger_{\textrm{B},j}+\mathbf{R}_{\textrm{dm},j}\bracket{k,l},\\
\mathbf{R}_{\textrm{dm},j}\bracket{k,l}&=\sum_{g\neq j}\mathbf{H}_{\textrm{B},j}\PBg\mathbf{P}^{\dagger}_{\mathrm{d},g}\bracket{k}\mathbf{H}^\dagger_{\textrm{B},j},\\ 
\textrm{and }
\mathbf{R}_{\mathrm{u},j}\bracket{k,l}&=\sum_{i=1}^{\mathit{I}}\mathbf{H}_{i,j}\PiB\PiBH\mathbf{H}^\dagger_{i,j},
\end{align} 
respectively. 

\color{black}Denote the cross-correlation function between $p_{\textrm{T}}\bracket{n}$ and $\phi^\dagger_{m\rr}\bracket{-n}$ by $\varphi_{\textrm{T},m\rr}\bracket{n}$. The UL signal through $\phi_{m\rr}^\dagger\bracket{-\paren{n+G}}$ and sampled at the symbol period $l$ of the $\ith{k}$ frame is
\begin{align}
    \mathbf{v}_{\textrm{u},i}\bracket{k,l}&=\sum_{i=1}^{I}\mathbf{H}_{i,\textrm{B}}\mathbf{x}_{\mathrm{u},i}\bracket{n}\ast \phi^\dagger_{m\rr}\bracket{-n-G}|_{n=kN+l}\nonumber\\
    &=\sum_{i=1}^{I}\mathbf{H}_{i,\textrm{B}}\mathbf{s}_{\mathrm{u},i}\bracket{k,l}\varphi_{\textrm{T},m\rr}\bracket{-G}.
\end{align}
Similar to the discussion on $v^{\paren{n_\target}}_{\textrm{r},n\rr}\bracket{k}$ in Section~\ref{sec: radar receive only},  we do not consider $\mathbf{v}_{\textrm{u},i}\bracket{k,l}$ for the derivation of system model and the CWSM algorithm. The same principle is applied to the DL signal model.  

\color{black}
\subsection{Radar Signals at Communications Rxs}
\label{subsubsec: radar-comm-Rx}
The signals radiated by $M\rr$ radar Txs and received by the BS Rx and the $j$-th DL UE are 
\begin{flalign}
\widetilde{\mathbf{y}}_{\mathrm{r},\mathrm{B}}\paren{t}&=\sum_{m\rr=1}^{M\rr}
s_{m\rr}\paren{t}\ast \mathbf{g}_{m\rr,\mathrm{B}}\paren{t,\tau}\nonumber\\
&\approx\sum_{m\rr=1}^{M\rr}\sum_{k=0}^{K-1}\mathbf{h}_{m\rr,\mathrm{B}}e^{j2\pi f_{m\rr\mathrm{B}}kT_{\mathrm{r}}} a_{m_\mathrm{r},k}\phi_{m_\mathrm{r}}\paren{t-kT\rr-GT_{\mathrm{p}}-\tau_{m\rr,\mathrm{B}}}\nonumber\\
\text{and }\widetilde{\mathbf{y}}_{\mathrm{r},j}\paren{t}&=\sum_{m\rr=1}^{M\rr}\mathbf{y}_{m\rr,j}\paren{t}=\sum_{m\rr=1}^{M\rr}
s_{m\rr}\paren{t}\ast \mathbf{g}_{m\rr,j}\paren{t,\tau}\nonumber\\
&\approx\sum_{m\rr=1}^{M\rr}\sum_{k=0}^{K-1}\mathbf{h}_{m\rr,j}e^{j2\pi f_{m\rr,j}kT_{\mathrm{r}}}a_{m_\mathrm{r},k}\phi_{m_\mathrm{r}}\paren{t-kT\rr-GT_{\mathrm{p}}-\tau_{m\rr,j}},\nonumber
\end{flalign}\normalsize
respectively. Discretizing $\tau_{m\rr,\mathrm{B}}$ and $\tau_{m\rr,j}$ and assuming narrowband signals, we have $\lfloor\sfrac{\tau_{m\rr,\mathrm{B}}}{T_\mathrm{p}}\rfloor=n_{\mathrm{rB}}\in\mathbb{Z}_{+}\paren{N}$ and $\lfloor\sfrac{\tau_{m\rr,j}}{T_\mathrm{p}}\rfloor=n_{\textrm{rd}}\in\mathbb{Z}_{+}\paren{N}$ for all $m\rr$ and $j$. The respective discrete-time sampled signals 
are $\widetilde{\mathbf{y}}_{\mathrm{rB}}\bracket{n}=\mathbf{y}_{\mathrm{rB}}\paren{nT_{\mathrm{p}}}$ and $\mathbf{y}_{\mathrm{r},j}\bracket{n}=\mathbf{y}_{\mathrm{r},j}\paren{nT_{\mathrm{p}}}$. 
After estimating the Doppler shifts $f_{m\rr,\B}$ and $f_{m\rr,j}$, the radar signal components at the outputs of the matched filters of BS Rx and $j$-th DL UE are, respectively, 
\begin{flalign}
\label{eq: radar_BS_1}
\mathbf{y}_{\mathrm{rB}}\bracket{n}&=\sum_{m\rr=1}^{M\rr}\widetilde{\mathbf{y}}_{\mathrm{rB}}\bracket{n}\ast \phi^\dagger_{m\rr}\bracket{-\paren{n+G}}e^{-j2\pi f_{m\rr,\B}kT\rr},
\end{flalign}
and
\begin{flalign}
\mathbf{y}_{\mathrm{r},j}\bracket{n}&=\sum_{m\rr=1}^{M\rr}\widetilde{{\mathbf{y}}}_{\mathrm{r},j}\bracket{n}\ast \phi^\dagger_{m\rr}\bracket{-\paren{n+G}}e^{-j2\pi f_{m\rr,j}kT\rr}\label{eq: radar_DL_1},
\end{flalign}
where \eqref{eq: radar_BS_1} and \eqref{eq: radar_DL_1} peak at $n=kN+n_{\mathrm{rB}}$ and $n=kN+n_{\textrm{rd}}$, respectively. 
Therefore, with 
\begin{align}
\mathbf{H}_{\textrm{rB}}=\bracket{\mathbf{h}_{1,\textrm{B}},\cdots,\mathbf{h}_{\mathit{M}\rr,\textrm{B}}}\in\mathbb{C}^{\mathit{M}\textrm{c}\times \mathit{M}\rr},
\end{align}
and
\begin{align}
\mathbf{H}_{\mathrm{r},j}\triangleq\bracket{\mathbf{h}_{\mathrm{1},j},\cdots,\mathbf{h}_{\mathit{M}\rr,j}},
\end{align}
the radar signals interfering with the $\ith{n_{\mathrm{rB}}}$ symbol period of the $\ith{k}$ UL frame at the BS Rx and $\ith{n_{\textrm{rd}}}$ symbol period of the $\ith{k}$ DL frame at the $j$-th DL UE are, respectively, 
\begin{flalign}
\mathbf{y}_{\mathrm{rB}}\bracket{k,n_{\mathrm{rB}}}\triangleq\mathbf{y}_{\mathrm{rtB}}\bracket{kN+n_{\mathrm{rB}}}=\mathbf{H}_{\mathrm{rB}}\mathbf{a}\bracket{k},  \label{eq: yrB}
\end{flalign}
and
\begin{flalign}
\mathbf{y}_{\mathrm{r},j}\bracket{k,n_{\textrm{rd}}}\triangleq\mathbf{y}_{\mathrm{r},j}\bracket{kN+n_{\textrm{rd}}}=\mathbf{H}_{\mathrm{r},j}\mathbf{a}\bracket{k}, \label{eq: yrj}
\end{flalign}
with CMs 
\begin{align}
\mathbf{R}_{\mathrm{rB}}\bracket{k,n_{\mathrm{rB}}}\in\mathbb{C}^{N_{\mathrm{c}}\times N_{\mathrm{c}}}=\mathbf{H}_{\mathrm{rB}}\mathbf{a}\bracket{k}\mathbf{a}^\dagger\bracket{k}\mathbf{H}^\dagger_{\mathrm{rB}},
\end{align}
and 
\begin{align}
\mathbf{R}_{\mathrm{r},j}\bracket{k,n_{\textrm{rd}}}\in\mathbb{C}^{N^\textrm{d}_{j}\times N^\textrm{d}_{j}}=\mathbf{H}_{\mathrm{r},j}\mathbf{a}\bracket{k}\mathbf{a}^\dagger\bracket{k}\mathbf{H}^\dagger_{\mathrm{r},j}. 
\end{align}

The cross-correlation function between $p_{\textrm{T}}\bracket{n}$ and $\phi_{m\rr}\bracket{n}$ is $\varphi_{m\rr,\textrm{R}}\bracket{n}$. The radar signal components filtered by  $p_{\textrm{R}}\bracket{n}$ at the BS and $j$-th DL UE are 
\begin{align}
\mathbf{v}_{\textrm{r},\textrm{B}}\bracket{n}&=\widetilde{\mathbf{y}}_{\mathrm{rB}}\bracket{n}\ast p_{\textrm{R}}\bracket{n}|_{n=kN+n_{\textrm{rB}}}\nonumber\\
&=\sum_{m\rr=1}^{M\rr}\mathbf{h}^\top_{m\rr,\B}\mathbf{a}_{m\rr,k}\varphi_{m\rr,\textrm{R}}\bracket{-G},
\end{align}
and 
\begin{align}
\mathbf{v}_{\textrm{r},j}\bracket{n}&=\widetilde{y}_{\mathrm{rB}}\bracket{n}\ast p_{\textrm{R}}\bracket{n}|_{n=kN+n_{\textrm{rd}}}\nonumber\\
&=\sum_{m\rr=1}^{M\rr}\mathbf{h}^\top_{m\rr,j}\mathbf{a}_{m\rr,k}\varphi_{m\rr,\textrm{R}}\bracket{-G},
\end{align}
 for all $j$. Since $\varphi_{m\rr,\textrm{R}}\bracket{-G}$ is a constant coefficient for all $m\rr$ and do not carry any communications information, these components at the output of $p_{\textrm{R}}\bracket{n}$ are not useful for communications symbol extraction.

\subsection{Composite BS and DL received signals}
\begin{figure}[!t]
	\centering
	\includegraphics[width=0.80\columnwidth]{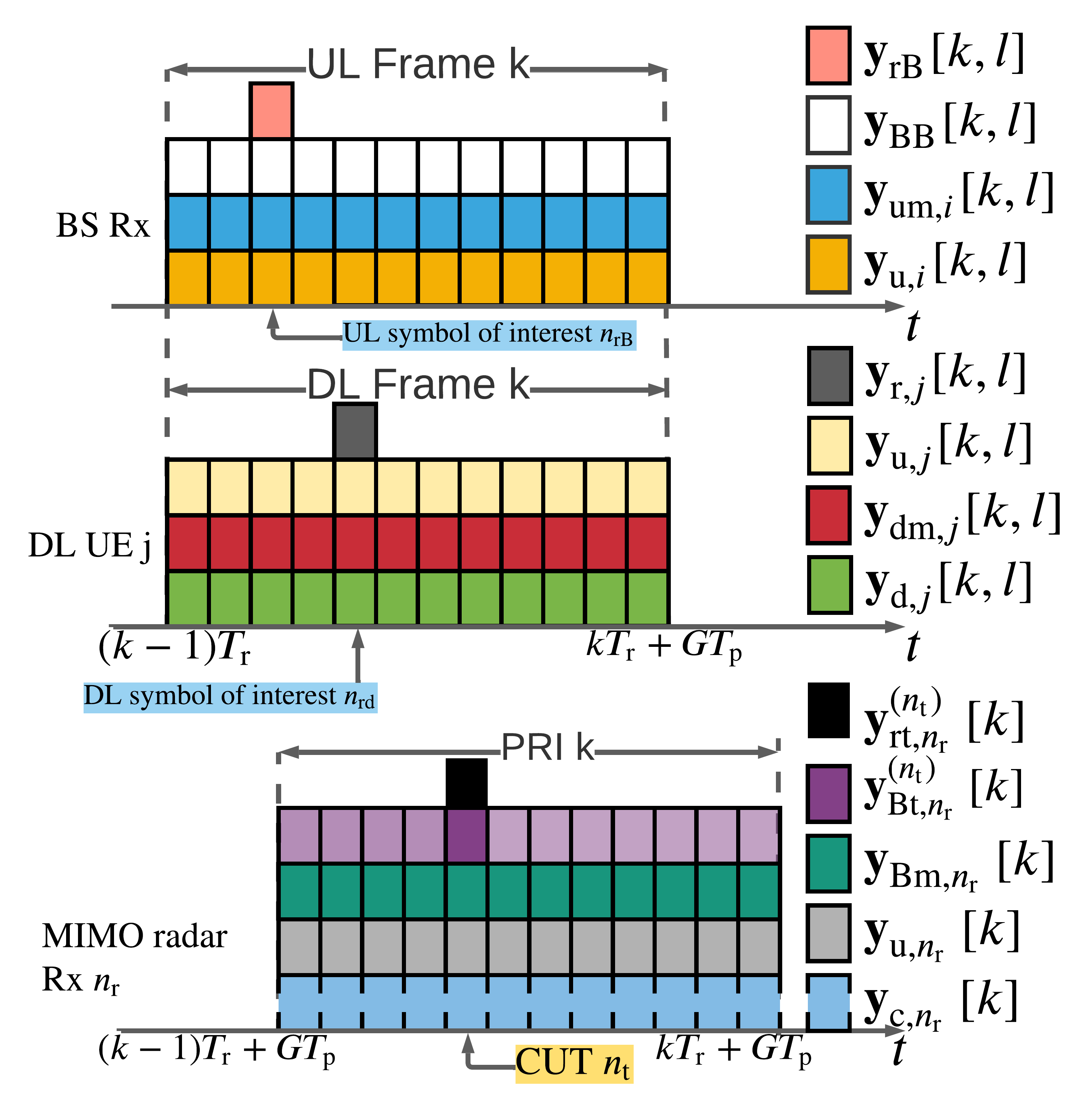}
	\caption{\textcolor{black}{The overlaid receive signal timing diagram during $\ith{k}$ radar PRI and $\ith{k}$ communications frame in the observation window; noise trails have been excluded. The purple bin with more opacity indicates the DL signal reflected from the target and observed in the radar CUT, i.e., $\mathbf{y}^{\paren{n_\target}}_{\textrm{Bt},n\rr}\bracket{k}$. Other more translucent purple bins indicate $\mathbf{y}^{\paren{n^\prime_\target}}_{\textrm{Bt},n\rr}\bracket{k}$ for $n^\prime_\target\neq n_\target$ (see \eqref{eq:Bt_range_cell}). } 
	} 
	\label{fig:ReceiveSignalModel}
\end{figure}

Denoting the CSCG noise vectors measured respectively at the BS Rx and the $\ith{j}$ DL UE as $\mathbf{z}_\textrm{B}\bracket{k,l}\sim\mathcal{CN}\paren{0,\sigma^2_{\textrm{B}}\mathbf{I}_{M\cc}}$ and $\mathbf{z}_{\textrm{d},j}\bracket{k,l}\sim\mathcal{CN}\paren{0,\sigma^2_{\mathrm{d},j}\mathbf{I}_{N^\mathrm{d}_{j}}}$, i.i.d in $k$ and $l$, the signal received at the BS Rx to decode $\mathbf{s}_{\textrm{u},i}\bracket{k,l}$ and the composite signal received by the $j$-th DL UE are, respectively, 
\begin{flalign}
\mathbf{y}_{\textrm{u},i}\bracket{k,l}&=\mathbf{y}_{i,\B}\bracket{k,l}+\mathbf{y}_{\textrm{um},i}\bracket{k,l}+\mathbf{y}_{\textrm{BB}}\bracket{k,l}+\mathbf{y}_{\textrm{rB}}\bracket{k,l}+\mathbf{z}_\textrm{B}\bracket{k,l},
\label{eq: UL_total_received}\\
\textrm{and }
\mathbf{y}_{\textrm{d},j}\bracket{k,l}&=\mathbf{y}_{\textrm{B},j}\bracket{k,l}+\mathbf{y}_{\mathrm{dm},j}\bracket{k,l}+\mathbf{y}_{\mathrm{u},j}\bracket{k,l}+\mathbf{y}_{\mathrm{r},j}\bracket{k,l}+\mathbf{z}_{\textrm{d},j}\bracket{k,l}. \label{eq: DL_total_received}
\end{flalign}
\normalsize
The CMs of $\mathbf{y}_{\textrm{u},i}\bracket{k,l}$ and $\mathbf{y}_{\textrm{d},j}\bracket{k,l}$ are, respectively, 
\begin{flalign}
\mathbf{R}_{\mathrm{u},i}\bracket{k,l}&=\mathbf{R}_{i,\B}\bracket{k,l}+ \mathbf{R}^\mathrm{in}_{\mathrm{u},i}\bracket{k,l},
\end{flalign}
and
\begin{flalign}
\mathbf{R}_{\mathrm{d},j}\bracket{k,l}&=\mathbf{R}_{\mathrm{B,j}}\bracket{k,l}+\mathbf{R}^\mathrm{in}_{\mathrm{d},j}\bracket{k,l},
\end{flalign}
where 
\begin{align}
\mathbf{R}^\mathrm{in}_{\mathrm{u},i}\bracket{k,l}=\mathbf{R}_{\textrm{um}, i}\bracket{k,l}+\mathbf{R}_{\mathrm{BB}}\bracket{k,l}+\mathbf{R}_{\textrm{rB}}\bracket{k,l}+\sigma^2_{\textrm{B}}\mathbf{I}_{\mathit{M}\cc},
\end{align}
and
\begin{align}
\mathbf{R}^{\textrm{in}}_{\textrm{d},j}\bracket{k,l}=\mathbf{R}_{\mathrm{dm},j}\bracket{k,l}+\mathbf{R}_{\mathrm{u,}j}\bracket{k,l}+\mathbf{R}_{\mathrm{r},j}\bracket{k,l}+\sigma^2_j\mathbf{I}_{N^{\textrm{d}}_{j}},
\end{align}
denote the interference-plus-noise CMs associated with \eqref{eq: DL_total_received} and \eqref{eq: UL_total_received}, respectively.
Note from \eqref{eq: yrB} and \eqref{eq: yrj} that $\mathbf{y}_{\mathrm{rB}}\bracket{k,l}=\mathbf{0}$ for $l\neq n_{\mathrm{rB}}$ and $\mathbf{y}_{\mathrm{r},j}\bracket{k,l}=\mathbf{0}$ for $l\neq n_{\textrm{rd}}$. \color{black}Hence, precoders of IBFD communications are based on the $\ith{n_{\mathrm{rB}}}$ symbol period of $K$ UL frames and the $\ith{n_{\textrm{rd}}}$ symbol period of $K$ DL frames, where $n_{\mathrm{rB}}$ and $n_{\textrm{rd}}$ are the symbol indices of UL and DL, respectively. 
\figurename{~\ref{fig:ReceiveSignalModel}} illustrates the composite receive signals of BS Rx, $j$-th DL UE, and $n\rr$-th radar Rx.

\section{CWSM Maximization}
\label{sec: formulation}
We now define the LRFs for the MIMO radar and the IBFD MU-MIMO communications system Rxs before introducing the MI-based co-design metric CWSM. Denote the LRF at the $\ith{n\rr}$ radar as $\mathbf{U}_{\mathrm{r},n\rr}=\bracket{\mathbf{u}_{\mathrm{r},n\rr}\bracket{0},\cdots,\mathbf{u}_{\mathrm{r},n\rr}\bracket{K-1}}\in\mathbb{C}^{KM\times\mathrm{\mathit{K}}}$. This LRF's output is 
\begin{align}
\widetilde{\mathbf{y}}_{\textrm{r},n\rr}&=\widetilde{\mathbf{y}}_{\mathrm{t},n\rr}+\widetilde{\mathbf{y}}^{\mathrm{in}}_{\mathrm{r},n\rr}=\mathbf{U}\rnr\mathbf{S}_{\mathrm{t},n\rr}\mathbf{h}_{\mathrm{t},n\rr}+\mathbf{U}\rnr\mathbf{y}^{\mathrm{in}}_{\mathrm{r},n\rr}
\end{align}
where $\mathbf{h}_{\mathrm{t},n\rr}\sim\mathcal{CN}\paren{\mathbf{0},\boldsymbol{\Sigma}_{\target,n\rr}}$ contains the target information and $\widetilde{\mathbf{y}}_{\mathrm{r},n\rr}\sim\mathcal{CN}\paren{\mathbf{0},\mathbf{U}\rnr\paren{\mathbf{R}_{\mathrm{t},n\rr}+\mathbf{R}^{\mathrm{in}}_{\mathrm{r},n\rr}}\mathbf{U}^\dagger\rnr}$. Hence, using the chain rule, the MI between $\widetilde{\mathbf{y}}_{\mathrm{r},n\rr}$ and 
$\mathbf{h}_{\mathrm{t},n\rr}$ is
\cite{Colornoise_waveform,Jammer_game}
\begin{align}
\mathit{I}\rnr&\triangleq\mathit{I}\paren{\widetilde{\mathbf{y}}\rnr;\mathbf{h}_{\mathrm{t},n\rr}}=\mathit{H}\paren{\widetilde{\mathbf{y}}\rnr}-\mathit{H}\paren{\widetilde{\mathbf{y}}\rnr|\mathbf{h}_{\mathrm{t},n\rr}}\nonumber\\
&=\mathit{H}\paren{\widetilde{\mathbf{y}}\rnr}-\mathit{H}\paren{\widetilde{\mathbf{y}}_{\textrm{t},n\rr}|\mathbf{h}_{\mathrm{t},n\rr}}-\mathit{H}\paren{\widetilde{\mathbf{y}}^{\mathrm{in}}_{\mathrm{r},n\rr}|\mathbf{h}_{\mathrm{t},n\rr}}\nonumber\\
&=\mathit{H}\paren{\widetilde{\mathbf{y}}\rnr}-\mathit{H}\paren{\widetilde{\mathbf{y}}^{\mathrm{in}}_{\mathrm{r},n\rr}},
\end{align}
where $\mathit{H}\paren{\widetilde{\mathbf{y}}_{\textrm{t},n\rr}|\mathbf{h}_{\mathrm{t},n\rr}}$ vanishes because $\widetilde{\mathbf{y}}_{\textrm{t},n\rr}$ depends on $\mathbf{h}_{\mathrm{t},n\rr}$; and $\mathit{H}\paren{\widetilde{\mathbf{y}}^{\mathrm{in}}_{\mathrm{r},n\rr}|\mathbf{h}_{\mathrm{t},n\rr}}$ reduces to $\mathit{H}\paren{\widetilde{\mathbf{y}}^{\mathrm{in}}_{\mathrm{r},n\rr}}$ because $\widetilde{\mathbf{y}}^{\textrm{in}}_{\textrm{r},n\rr}$ and $\mathbf{h}_{\textrm{t},n\rr}$ are mutually independent. 
The conditional differential entropy with the Gaussian noise \cite{Colornoise_waveform} leads to  
\begin{align}
\mathit{H}\paren{\widetilde{\mathbf{y}}\rnr|\mathbf{A}}=\varrho+\log\left|\mathbf{U}\rnr\mathbf{R}_{\mathrm{r},n\rr}\mathbf{U}^\dagger\rnr \right|
\end{align}
and
\begin{align}
\mathit{H}\paren{\widetilde{\mathbf{y}}^{\mathrm{in}}_{\mathrm{r},n\rr}|\mathbf{A}}=\varrho+\log\left|\mathbf{U}\rnr\mathbf{R}^{\mathrm{in}}_{\mathrm{r},n\rr}\mathbf{U}^\dagger\rnr \right|,
\end{align}
where the constant $\varrho=K\log\paren{\pi}+K$. This gives
\begin{align}\label{radarmi}
	\mathit{I}\rnr&=\log\frac{\left|\mathbf{U}\rnr\paren{\mathbf{R}_{\mathrm{t},n\rr}+\mathbf{R}^{\mathrm{in}}_{\mathrm{r},n\rr}}\mathbf{U}^\dagger\rnr \right|}{\left|\mathbf{U}\rnr\mathbf{R}^{\mathrm{in}}_{\mathrm{r},n\rr}\mathbf{U}^\dagger\rnr \right|}\nonumber\\
	&=\log\left\lvert\mathbf{I}+\mathbf{U}\rnr\mathbf{R}_{\textrm{t},n\rr}\mathbf{U}^\dagger\rnr\paren{\mathbf{U}\rnr\mathbf{R}^{\mathrm{in}}_{\mathrm{r},n\rr}\mathbf{U}^\dagger\rnr}^{-1}\right\rvert.
\end{align}

The LRFs deployed at the BS to decode the $\ith{i}$ UL UE and $\ith{j}$ DL UE during $\ith{k}$ frame of the observation window are  $\UiB\in\mathbb{C}^{\mathit{D}^{\textrm{u}}_{i}\times N\cc}$ and $\UBj\in\mathbb{C}^{\mathit{D}^{\textrm{d}}_{j}\times N^{\textrm{d}}_{j}}$, respectively. The outputs of $\UiB$ and $\UBj$ are $\widetilde{\mathbf{y}}_{\mathrm{u},i}\bracket{k,l}=\UiB\mathbf{y}_{\mathrm{u},i}\bracket{k,l}$ and $\widetilde{\mathbf{y}}_{\mathrm{d},j}\bracket{k,l}=\UBj\mathbf{y}_{\mathrm{d},j}\bracket{k,l}$; these signals follow the distributions $\mathcal{CN}\paren{\mathbf{0},\UiB\mathbf{R}_{\mathrm{u},i}\bracket{k,l}\UiBH}$ and $\mathcal{CN}\paren{\mathbf{0},\UBj\mathbf{R}_{\mathrm{d},j}\bracket{k,l}\UBjH}$, respectively. 

Using the common assumption of Gaussianity on symbol vectors, i.e. $\mathbf{s}_{\mathrm{u},i}\bracket{k,l}\sim\mathcal{CN}\paren{\mathbf{0},\PiB\PiBH}$ and $\mathbf{s}_{\mathrm{d},j}\bracket{k,l}\sim\mathcal{CN}\paren{\mathbf{0},\PBj\PBjH}$, the MIs between $\widetilde{\mathbf{y}}_{\mathrm{u},i}\bracket{k,l}$ and $\mathbf{s}_{\mathrm{u},i}\bracket{k,l}$ as well as $\widetilde{\mathbf{y}}_{\mathrm{d},j}\bracket{k,l}$ and $\mathbf{s}_{\mathrm{d},j}\bracket{k,l}$ are
\begin{flalign}
\mathit{I}^{\textrm{u}}_{i}\bracket{k,l}&\triangleq \mathit{I}\paren{\mathbf{s}_{\mathrm{u},i}\bracket{k,l};\widetilde{\mathbf{y}}_{\mathrm{u},i}\bracket{k,l}}
=\log\left|\mathbf{I}+\UiB\mathbf{R}_{i,\B}\bracket{k,l}\UiBH\paren{\UiB\mathbf{R}^{\mathrm{in}}_{\mathrm{u},i}\bracket{k,l}\UiBH}^{-1}\right|,
\end{flalign}
and
\begin{flalign}
\mathit{I}^{\textrm{d}}_{j}\bracket{k,l}&\triangleq I\paren{\mathbf{s}_{\mathrm{B},j}\bracket{k,l};\widetilde{\mathbf{y}}_{\mathrm{d},j}\bracket{k,l}}=\label{DLmutual} =\log\left|\mathbf{I}+\UBj\mathbf{R}_{\textrm{B},j}\bracket{k,l}\UBjH\paren{\UBj\mathbf{R}^{\mathrm{in}}_{\mathrm{d},j}\bracket{k,l}\UBjH}^{-1}\right|,
\end{flalign}
respectively. Recall from Section~\ref{subsubsec: radar-comm-Rx} 
that we evaluate FD communications based on the symbols-of-interest, i.e., $l=n_{\mathrm{rB}}$ for each UL frame and $l=n_{\textrm{rd}}$ for each DL frame in the observation window. The metric CWSM is a weighted sum of communications' MIs related to the symbol periods of interest\footnote{Hereafter, for simplicity, we drop symbol index $l=n_{\textrm{rB}}$ ($l=n_{\textrm{rd}}$) for UL (DL) related terms.} and $I_{\mathrm{r},n\rr}$, i.e., 
\begin{flalign}
\label{objectfunction1}
\mathit{I}_{\textrm{CWSM}}&=\sum_{n\rr=1}^{N\rr}\alpha^\textrm{r}_{n\rr} \mathit{I}^{\textrm{r}}_{n\rr}+\sum_{k=0}^{K-1}\bracket{\sum_{i=1}^\mathit{I}\alpha^\textrm{u}_i\mathit{I}^{\textrm{u}}_{i}\bracket{k}+\sum_{j=1}^\mathit{J}\alpha^\textrm{d}_j\mathit{I}^{\textrm{d}}_{j}\bracket{k}},
\end{flalign}
where $\alpha^\textrm{r}_{n\rr}$, $\alpha^\textrm{u}_i$, and $\alpha^\textrm{d}_j$ are pre-defined weights assigned to the MIMO $n\rr$-th radar Rx , $\ith{i}$ UL UE and $\ith{j}$ DL UE, respectively, for all $n\rr$, $i$, and $j$; the weights are determined by the system priority and specific applications. For example, for FD communications, the weights are based on available buffer capacities of BS and UEs\cite{MSE_FD,FD_WMMSE}. For the joint radar-communications, one can assign larger (smaller) weights to $\alpha^{\mathrm{r}}_{n_\mathrm{r}}$ with the presence (absence) of targets 
\cite{Liu2018Gloabalsip}.

Denote the sets of the precoders and the LRFs as $\braces{\mathbf{P}}\triangleq\braces{\PiB,\PBj | i\in\mathbb{Z}_+\braces{I}, j\in\mathbb{Z}_+\braces{J}, k\in\mathbb{Z}_+\braces{K}}$
and
$\braces{\mathbf{U}}\triangleq\left\lbrace\UiB, \UBj, \mathbf{U}\rnr | i\in\mathbb{Z}_+\braces{I}, j\in\mathbb{Z}_+\braces{J},\right.$ $\left.k\in\mathbb{Z}_+\braces{K}, n\rr\in \mathbb{Z}_+\braces{N\rr}\right\rbrace$. 
The transmission powers that occurred to the BS and the $i$-th UL UE at the $\ith{k}$ frame are
\begin{align}
\label{eq: DL_power}
P_{\mathrm{d}}\bracket{k}=\sum_{j=1}^{J}P_{\mathrm{d},j}\bracket{k}=\sum_{j=1}^{J}\trace\braces{\PBj\PBjH},
\end{align}
and 
\begin{align}
P_{\mathrm{u},i}\bracket{k}=\trace\braces{\PiB\PiBH}, \label{eq: UL_power}
\end{align}
which are upper bounded by the maximum DL and UL powers $P_\B$ and $P_{\mathrm{U}}$, respectively. The achievable rates for the $i$-th UL UE and the $j$-th DL UE in the $\ith{k}$ frame, $R_{\mathrm{u},i}\bracket{k}$ and $R_{\mathrm{d},j}\bracket{k}$ are lower bounded by the least acceptable achievable rates to quantify the QoS of the UL and DL, $\mathit{R}_{\textrm{UL}}$ and $\mathit{R}_{\textrm{DL}}$, respectively. The CWSM optimization to jointly design precoders $\braces{\mathbf{P}}$, radar code $\mathbf{A}$, and LRFs $\braces{\mathbf{U}}$ is 
\begin{subequations}\label{jointop}\begin{align}
	\underset{{\braces{\mathbf{P}},\braces{\mathbf{U}},
			\mathbf{A}}}{\text{maximize}}\;& \mathit{I}_{\textrm{CWSM}}\paren{\braces{\mathbf{U}},\braces{\mathbf{P}},\mathbf{A}}  \\
	\text{subject to}\; & P_{\mathrm{d}}\bracket{k}\leq             
                            P_\textrm{B},\label{DL_power}\\
                            &P_{\mathrm{u},i}\bracket{k}\leq P_\textrm{U}, \\*
	                       &\mathit{R}_{\textrm{u},i}\bracket{k}\geq\mathit{R}_{\textrm{UL}}, \\
                           &\mathit{R}_{\textrm{d},j}\bracket{k}\geq \mathit{R}_{\textrm{DL}},\label{DLrate}\\
	                   &\lVert\mathbf{a}_{m\rr}\rVert^2 =P_{\textrm{r},m\rr},\; \label{constraint:radarpower}\\*
	                    &\frac{\mathrm{\mathit{K}}\max_{k=1,\cdots,  K}\lvert\mathbf{a}_{m\rr}\bracket{k}\rvert^2}{P_{\mathrm{r},m\rr}}\leq\mathrm{\gamma}_{m\rr},\; \forall\; i,j,k,m\rr,\label{constraint:radarpar}
\end{align}
\end{subequations}
where constraints \eqref{constraint:radarpower} and \eqref{constraint:radarpar}  are determined by the transmit power and PAR of the $\ith{m\rr}$ MIMO radar Tx, respectively. Note that PAR constraint is applied column-wise to the code matrix $\mathbf{A}$ because the Txs of a statistical MIMO radar are widely distributed. When $\gamma_{m\rr}$ = 1, PAR constraint is reduced to constant modulus constraint. 

We remark that the radar system in our problem is a statistical MIMO radar with widely distributed antennas instead of the colocated MIMO found in most existing works. The spatial diversity of the statistical MIMO radar enables all Tx-Rx pairs to observe statistically independent RCSs of the same target and hence improve its detection. On the other hand, the RCS is identical for all (closely-spaced) Txs and Rxs of a colocated MIMO radar thereby rendering any exploitation of spatial diversity ineffective \cite{haimovich2008mimo}. We utilize $I_{\mathrm{CWSM}}$ as a common information-theoretic design metric for both radar and communications. Some prior MRMC studies have utilized similar MI-based metrics, but were limited to only colocated MIMO radars \cite{alaee2020information,dokhanchi2020multi}. 
The constraints in \eqref{jointop} are also more comprehensive when compared with similar problems in the prior art focused on, again, colocated MRMC. The inclusion of power budget, QoS, and PAR in our problem makes the design more practical than previous colocated MRMC studies that have utilized only a subset of these constraints \cite{alaee2020information,dokhanchi2020multi, Singh2020FDRadar}. The following companion paper (Part II) \cite{liu2024codesigningpart2} develops the BCD-AP algorithm to solve the problem \eqref{jointop}.


\section{Numerical Experiments}
\label{sec:numerical}
We validated our spectral co-design approach through extensive numerical experiments. Throughout this section, we assume the noise variances $\sigma^2_{\textrm{r}}=\sigma^2_{\B}=\sigma^2_{\textrm{d}}=0.001$. We assume unit small scale fading channel gains, namely, the elements of $\mathbf{H}_{\textrm{B},j}$, $\mathbf{H}_{i,\textrm{B}}$, $\mathbf{H}_{i,j}$, $\boldsymbol{\alpha}_{\textrm{Bm},n\rr}$, and $\boldsymbol{\alpha}_{i,n\rr}$ are drawn from $\mathcal{CN}\paren{0,1}$. We model the self-interfering channel $\HBB$ as $\mathcal{CN}\paren{\sqrt{\frac{\sigma^2_{\mathrm{SI}}K_{\B}}{1+K_{\B}}}\widehat{\mathbf{H}}_{\textrm{BB}},\frac{\sigma^2_{\mathrm{SI}}}{1+K_{\B}}\mathbf{I}_{N\cc}\otimes\mathbf{I}_{M\cc}}$, where $\sigma^2_{\mathrm{SI}}$ is the SI attenuation coefficient that characterizes the effectiveness of SI cancellation \cite{MSE_FD}, the Rician factor $K_{\B}=1$, and $\widehat{\mathbf{H}}_{\textrm{BB}}\in\mathbb{C}^{N\cc\times M\cc}$ is an all-one matrix \cite{FD_WMMSE}. Define the signal-to-noise ratios (SNRs) associated with the MIMO radar, DL, and UL as $\mathrm{SNR}_{\textrm{r}}=\sfrac{P_{\textrm{r},m\rr}}{\sigma^2_{\textrm{r}}}$, $\mathrm{SNR}_{\textrm{DL}}=\sfrac{P_{\textrm{B}}}{\sigma^2_{\textrm{d}}}$, and $\mathrm{SNR}_{\textrm{UL}}=\sfrac{P_{\textrm{u},i}}{\sigma^2_{\textrm{\B}}}$ \cite{Luo2011IterativeWMMSE}. The clutter power $\sigma^2_{\textrm{c}}=\sigma^2_{m\rr\textrm{c}n\rr}$ for all $m\rr$ and $n\rr$ and clutter-to-noise ratio (CNR) is $\mathrm{CNR}=\sfrac{\sigma^2_{\textrm{c}}}{\sigma^2_0}$. Then, together with the direct path components, they are received at the IBFD communications Rxs. We model $\mathbf{h}_{m\rr,\textrm{B}}$ and $\mathbf{h}_{m\rr, j}$ as $\mathcal{CN}\paren{\sqrt{\frac{1}{\kappa+1}}\boldsymbol{\mu}_{m\rr,\textrm{B}},\frac{\eta^2_{\mathrm{m\rr},\textrm{B}}}{\kappa+1}\mathbf{I}_{N\cc}}$, and $\mathcal{CN}\paren{\sqrt{\frac{1}{\kappa+1}}\boldsymbol{\mu}_{m\rr,j},\frac{\eta^2_{\mathrm{m\rr},j}}{\kappa+1}\mathbf{I}_{N^{\textrm{d}}_{j}}}$, where $\kappa=1$, $\boldsymbol{\mu}_{m\rr,\B}=0.1\mathbf{1}_{N\cc}$, $\boldsymbol{\mu}_{m\rr,j}=0.05\mathbf{1}_{N^{\textrm{d}}_{j}}$,  $\eta^2_{m\rr,\B}=0.3$, $\eta^2_{m\rr,j}=0.5$.

Unless otherwise stated, we use the following parameter values: number of radar Txs and Rxs: $\mathit{M}\rr=N\rr=4$; number of communications Tx and Rx antennas:  $\mathit{M}\cc=N\cc=4$; $\mathit{I}=\mathit{J}=2$; $N^{\textrm{u}}_{i}=\mathrm{d}^{\textrm{u}}_i=N^{\textrm{d}}_{j}=\mathrm{d}^{\textrm{d}}_j=2$,  for all $\braces{i,j}$; $\mathrm{SNR}_{\textrm{DL}}=\mathrm{SNR}_{\textrm{UL}}=10$ dB; $\sigma^2_{\mathrm{SI}}=0$ dB; \textcolor{black}{$\mathrm{CNR}=20$ dB}; radar PAR $\gamma_{m\rr}=3$ dB; number of communications frames or radar PRIs $\mathit{K}=8$; number of symbols in each frame or range cells in each radar PRI $N=32$; radar CUT index $n_\target=4$; UL (DL) indices of interest $n_{\textrm{rB}}=2$ ($n_{\textrm{r,d}}=3$); QoS of UL (DL): $\mathit{R}_{\textrm{u}}=\log_2(1+\frac{\mathrm{SNR}_{\textrm{UL}}}{M\rr*\mathrm{SNR}_{\textrm{r}}+\mathrm{SNR}_{\textrm{DL}}+(I-1)*\mathrm{SNR}_{\textrm{UL}}})$ bits/s/Hz ($\mathit{R}_{\textrm{d}}=\log_2(1+\frac{\sfrac{\mathrm{SNR}_{\textrm{DL}}}{J}}{M\rr*\mathrm{SNR}_{\textrm{r}}+\mathrm{SNR}_{\textrm{DL}}*\sfrac{(J-1)}{J}+I*\mathrm{SNR}_{\textrm{UL}}})$ bits/s/Hz); 
The normalized Doppler shifts $f_{m\rr\textrm{t}n\rr}T\rr$ and  $f_{\textrm{Bt}n\rr}T\rr$ are uniformly distributed in $\bracket{0.05,0.325}$ for each channel realization \cite{NaghshTSP2017}. As detailed in the following companion paper (Part II) \cite{liu2024codesigningpart2}, the numbers of iterations for the subgradient, weighted minimum mean-squared-error (WMMSE)-MRMC, and BCD-AP MRMC algorithms are $t_{\textrm{u,max}}=t_{\textrm{d,max}}=200$, $\mathrm{\iota}_{\textrm{max}}=1$, $\mathrm{\ell}_{\textrm{max}}=2000$. We use uniform weights  $\alpha^\textrm{u}_{i}= \alpha^{\textrm{d}}_{j}=\alpha^\textrm{r}_{n\rr}=$ $\frac{1}{\paren{\mathit{I}+\mathit{J}+\mathit{N}\rr}}$ for all $\braces{n\rr,i,j}$.

\subsection{Radar Detection Performance}
\begin{figure}[t]
\centering
\includegraphics[width=1.0\columnwidth]{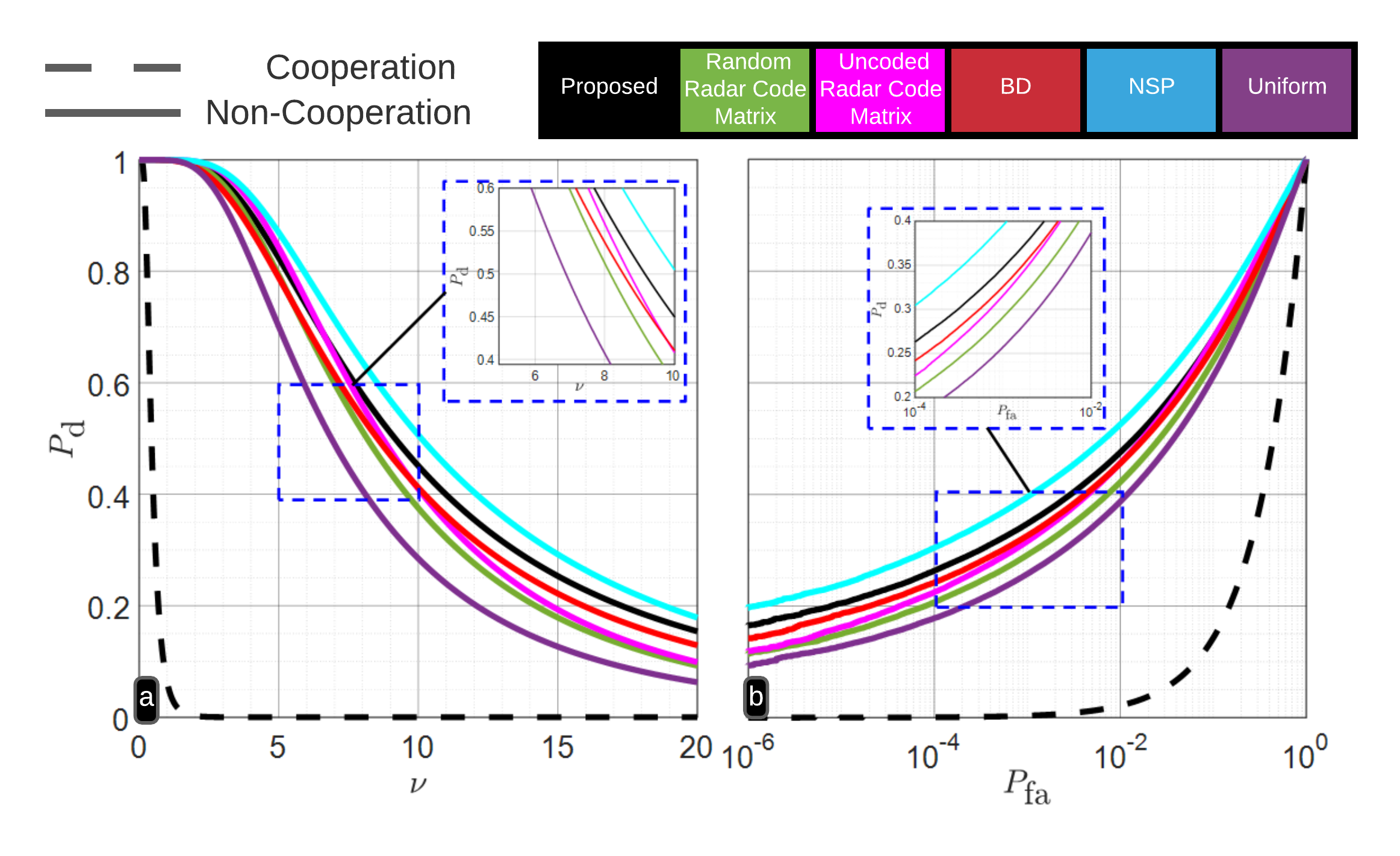}
\caption{Target detection performance of the co-designed system compared with other radar codes and cooperation schemes using the NP detector. (a) $P_{\textrm{d}}$ versus $\nu$ (b) ROC of the NP detector. }
\label{fig: NPdetector}
\end{figure}
We investigate the detection performance of the statistical MIMO radar using the designed code matrix $\mathbf{A}$. 
Consider the binary hypothesis testing formulation for target detection
\begin{equation}
\label{eq: hypothesis1}
\begin{cases}
\mathcal{H}_{\mathrm{0}}: & \mathbf{y}_{\textrm{r}} = \mathbf{y}^{\textrm{in}}_{\textrm{r}}
\\
\mathcal{H}_{\mathrm{1}}: & \mathbf{y}_{\mathrm{r}} = \mathbf{y}_{\textrm{tr}}+ \mathbf{y}^{\textrm{in}}_{\textrm{r}}.
\end{cases}
\end{equation}
\normalsize
With $\overline{\mathbf{y}}_{\textrm{r},n\rr} = \mathbf{R}^{-\sfrac{1}{2}}_{\textrm{in},n\rr}\mathbf{y}_{\textrm{r},n\rr}\in\mathbb{C}^{K}$ and its CM $\mathbf{G}_{n\rr}=\mathbf{R}^{-\sfrac{1}{2}}_{\textrm{in},n\rr}\mathbf{R}_{\textrm{t},n\rr}\mathbf{R}^{-\sfrac{1}{2}}_{\textrm{in},n\rr}$, one can rewrite \eqref{eq: hypothesis1} as  
\begin{equation}
	\label{eq: hypothesis2}
\begin{cases}
\mathcal{H}_{\mathrm{0}}: & \overline{\mathbf{y}}_{\mathrm{r}}\sim\mathcal{CN}\paren{\mathbf{0},\mathbf{I}_{Kn\rr}}
\\
\mathcal{H}_{\mathrm{1}}: & \overline{\mathbf{y}}_{\mathrm{r}}\sim\mathcal{CN}\paren{\mathbf{0},\mathbf{I}_{KN\rr}+\mathbf{G}},
\end{cases}
\end{equation}\normalsize
where the block diagonal matrix $\mathbf{G}\in\mathbb{C}^{KN\rr\times KN\rr}=\oplus_{n\rr=1}^{N\rr}\mathbf{G}_{n\rr}$. The eigendecomposition of $\mathbf{G}_{n\rr}$ is $\mathbf{G}_{n\rr}=\mathbf{V}_{n\rr}\mathbf{\Lambda}_{n\rr}\mathbf{V}^\dagger_{n\rr}$, where the columns of $\mathbf{V}_{n\rr}$ and the diagonal entries of $\mathbf{\Lambda}_{n\rr}\triangleq\diag\bracket{\delta_{1,n\rr},\cdots,\delta_{\mathit{K},n\rr}}$ are, respectively, the eigenvectors and eigenvalues of $\mathbf{G}_{n\rr}$, with the $\ith{k}$ eigenvalue $\delta_{k,n\rr}$. Using the Woodbury matrix identity and the eigendecomposition of $\mathbf{G}_{n\rr}$, the test statistic is 
$T\paren{\overline{\mathbf{y}}}=\sum_{n\rr=1}^{N\rr}T\paren{\overline{\mathbf{y}}_{\textrm{r},n\rr}}=\sum_{n\rr=1}^{N\rr}\overline{\mathbf{y}}^\dagger_{\textrm{r},n\rr}\paren{\mathbf{I}-\paren{\mathbf{G}_{n\rr}+\mathbf{I}}^{-1}}\overline{\mathbf{y}}_{\textrm{r},n\rr}
=\sum_{n\rr=1}^{N\rr}\overline{\mathbf{y}}^\dagger_{\textrm{r},n\rr}\mathbf{V}_{n\rr}\paren{\mathbf{\Lambda}^{-1}+\mathbf{I}}^{-1}\mathbf{V}^\dagger_{n\rr}\overline{\mathbf{y}}_{\textrm{r},n\rr}$.

Denote $\widehat{\mathbf{y}}_{\textrm{r},n\rr}=\mathbf{V}^\dagger_{n\rr}\overline{\mathbf{y}}_{\textrm{r},n\rr}=\bracket{\widehat{y}_{n\rr}\bracket{0},\cdots,\widehat{y}_{n\rr}\bracket{K-1}}$. Then, the Neyman-Pearson (NP) detector is \cite{cover2006elements} 
\begin{equation}
\label{eq: NPdetector}
T\paren{\overline{\mathbf{y}}}=\sum_{n\rr=1}^{N\rr}\sum_{k=0}^{K-1}\frac{\delta_{k,n\rr}\lvert\widehat{y}_{n\rr}\bracket{k}\rvert^2}{1+\delta_{k,n\rr}}\underset{\mathrm{H}_2}{\overset{\mathrm{H}_1}{\gtrless}}\nu,
\end{equation}
where $\nu$ is a threshold chosen to ensure a given probability of false alarm. We performed Monte Carlo simulations to evaluate the probability of detection $P_{\textrm{d}}$ and Rx operating characteristic (ROC) (curve of $P_{\textrm{d}}$ versus probability of false alarm $P_{\textrm{fa}}$) of the NP detector. \figurename{~\ref{fig: NPdetector}}a and \ref{fig: NPdetector}b show $P_{\textrm{d}}$ w.r.t. $\nu$ and ROC, respectively, for various coding schemes and cooperation modes. Here, the presence or absence of cooperation indicates whether or not DL signals $\mathbf{y}_{\textrm{Bt},n\rr}$ are incorporated in $\mathbf{y}_{\textrm{t,}n\rr}$ for all $n\rr$. For $\ith{m\rr}$ radar Tx, the uncoded waveform is $\mathbf{a}_{m\rr}=\sqrt{\frac{P_{\textrm{r},m\rr}}{\mathit{K}}}\mathbf{1}_{\mathit{K}}$ and randomly coded waveform is  $\mathbf{a}_{m\rr}=\sqrt{\frac{P_{\textrm{r},m\rr}}{\mathit{K}}}\mathbf{u}_{m\rr}$, where $\braces{\mathbf{u}_{m\rr}}$ is a randomly generated unitary basis subject to the PAR constraint. 

We evaluated our proposed radar code design for cases when the IBFD MU-MIMO communications system uses conventional precoding strategies, e.g., uniform precoding for the UL \cite{MCMIMO_RadComm} and block diagonal (BD) and the null-space projection (NSP) methods for DL\cite{Multiuser}. We generated $10^{6}$ realizations of $\overline{\mathbf{y}}_{\textrm{r}}$ under hypothesis $\mathcal{H}_1$ to estimate $P_{\textrm{d}}$ and $\mathcal{H}_0$ to estimate $P_{\textrm{fa}}$ based on $\nu$ for each case with $\mathrm{SNR}_{\textrm{r}}=0$ dB. \figurename{~\ref{fig: NPdetector}}a and \ref{fig: NPdetector}b illustrate that our optimized radar code matrix outperforms the uncoded and random coding schemes, and that the cooperation between the radar and BS boosts the radar detection performance. For example, with $P_{\textrm{fa}}=10^{-3}$, the proposed algorithm yields about $9\%$ to $20\%$ improvement in $P_{\textrm{d}}$ over uncoded radar code matrix and random code with cooperation strategies, respectively. We also notice that the detection performance remains similar for different precoding strategies when the proposed radar code matrix is utilized. The NSP method demonstrates a better radar detection performance in that it projects the communications signals onto the null space of the channels between the BS and the radar Rxs, reducing the DL interference. However, this projection may not benefit the DL UEs, resulting in the performance degradation of the  DL achievable rates illustrated in the following example. 

\subsection{FD Communications Performance}
\label{subsec: fd_comm_eva}
\begin{figure}[t]
\centering
\includegraphics[width=1\columnwidth]{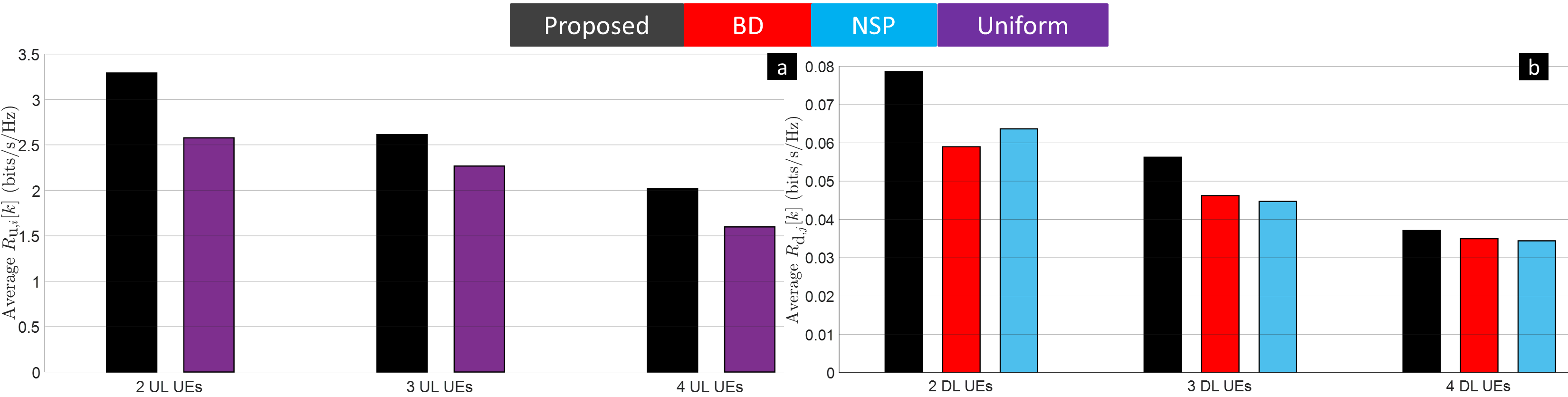}
\caption{Performance of the IBFD MU-MIMO communications system compared with varying numbers of (a) UL and (b) DL UEs.  }
\label{fig:fd_UE}
\end{figure}
We compared the IBFD MU-MIMO performance using the proposed precoders with existing precoding strategies for different numbers of UL/DL UEs in \figurename{~\ref{fig:fd_UE}}, where we plot average $R_{\textrm{u},i}\bracket{k}$ versus numbers of UL UEs in \figurename{~\ref{fig:fd_UE}}a with $\mathrm{SNR}_{\textrm{DL}}=\mathrm{SNR}_{\textrm{UL}}=10$ dB, and $R_{\textrm{d},j}\bracket{k}$ versus numbers of DL UEs in \figurename{~\ref{fig:fd_UE}}b with $\mathrm{SNR}_{\textrm{DL}}=0$ dB, $\mathrm{SNR}_{\textrm{UL}}=10$ dB. We also set $D^\textrm{u}_{i}=D^\textrm{d}_{j}=1$ to ensure that the spatial multiplexing is maintained. Both UL and DL communications demonstrate superior performances of the proposed precoding method over other benchmark ones.  

\section{Summary}
\label{sec:conclusion}
We proposed a spectral co-design framework for a statistical MIMO radar and an IBFD MU-MIMO communications system. Prior works primarily consider co-located MIMO radars, focus on co-existence solutions, and partially analyze MIMO communications. We take a wholesome view of the problem by jointly designing several essential aspects of such a co-design:  UL/DL precoders, MIMO radar code matrix, and LRFs for both systems. The radar codes generated by BCD-AP MRMC significantly increase $P_d$ over conventional coding schemes. We showed that cooperation between radar and DL signals is beneficial for target detection. The co-designed DL and radar are resilient to considerable UL interference. Similarly, using our optimized precoders and radar codes, the DL and UL rates remain stable as the CNR increases. In the following companion paper (Part II) \cite{liu2024codesigningpart2}, we develop the BCD-AP algorithm, provide theoretical guarantees, show its convergence, and examine joint radar-communications performance. The final companion paper (Part III) \cite{liu2024codesigningpart3} investigates the distributing co-phasing and multi-target tracking issues in a distributed MRMC.

	\bibliographystyle{elsarticle-num}
 \bibliography{SP_SI_Part1_v01}
	
\end{document}